\documentclass[aps,prb,twocolumn,superscriptaddress,longbibliography]{revtex4-1}

\usepackage{amsmath}
\usepackage{amsfonts}
\usepackage{amssymb}
\usepackage[colorlinks=true,citecolor=blue,linkcolor=red]{hyperref}
\usepackage{graphicx}
\usepackage{bbold}					% for \mathbb{1} as a nice unit matrix symbol
\usepackage[dvipsnames]{xcolor}

\usepackage{array,booktabs,ragged2e}	
\usepackage{soul}						% strike out text using \st{}
\usepackage{cancel}						% detto, but more clearly visible when reading
\usepackage{tabularx}					% for table in the CHS-model discussion
\usepackage{multirow}
\usepackage{hhline}

\usepackage[utf8]{inputenc}
\usepackage[T1]{fontenc}

\usepackage{adjustbox}

\usepackage{ulem}

\newcolumntype{P}[1]{>{\raggedleft\arraybackslash}p{#1}}
\newcolumntype{R}[1]{>{\centering\arraybackslash}p{#1}}

\begin{document}

%%%%%%%%%%%%%%%%%%%%%%%%%
%%% TITLE INFORMATION %%%
%%%%%%%%%%%%%%%%%%%%%%%%%

\title{MgTa$_2$N$_3$: A  reference Dirac semimetal}

\author{QuanSheng Wu}
\email{quansheng.wu@epfl.ch}
\affiliation{Institute of Physics, Ecole Polytechnique F\'{e}d\'{e}rale de Lausanne (EPFL), CH-1015 Lausanne, Switzerland}
\affiliation{National Centre for Computational Design and Discovery of Novel Materials MARVEL, Ecole Polytechnique F\'{e}d\'{e}rale de Lausanne (EPFL), CH-1015 Lausanne, Switzerland}
\affiliation{Institut f\"{u}r Theoretische Physik, ETH Z\"{u}rich, 8093 Z\"{u}rich, Switzerland}
\author{Christophe Piveteau}
\affiliation{Institut f\"{u}r Theoretische Physik, ETH Z\"{u}rich, 8093 Z\"{u}rich, Switzerland}
\author{Zhida Song}
\affiliation{Beijing National Laboratory for Condensed Matter Physics, and Institute of Physics, Chinese Academy of Science, Beijing 100190, China}
\author{Oleg V. Yazyev} 	
\email{oleg.yazyev@epfl.ch}
\affiliation{Institute of Physics, Ecole Polytechnique F\'{e}d\'{e}rale de Lausanne (EPFL), CH-1015 Lausanne, Switzerland}
\affiliation{National Centre for Computational Design and Discovery of Novel Materials MARVEL, Ecole Polytechnique F\'{e}d\'{e}rale de Lausanne (EPFL), CH-1015 Lausanne, Switzerland}

\date{\today}

\begin{abstract} 
We present a prediction of the Dirac semimetal (DSM) phase in MgTa$_2$N$_3$ based on first-principles calculations and symmetry analysis. In this material, the Fermi level is located exactly at the Dirac point without additional Fermi surface pockets. The band inversion associated with the Dirac cone involves the $d$ orbitals of two structurally inequivalent Ta atoms with octahedral and trigonal prismatic coordination spheres. We further show that the lattice symmetry breaking can realize topological phase transitions from the DSM phase to a triple nodal point semimetal, Weyl semimetal or topological insulator. The topologically protected surface states and the non-protected Fermi arc surface states are also studied.
\end{abstract} 

\maketitle

%introduction
%\section{Introduction}
 
Three-dimensional topological Dirac semimetals (DSMs)~\cite{PhysRevLett.108.140405,PhysRevB.85.195320, PhysRevB.88.125427} are materials realizing a novel state of quantum matter described by the massless Dirac equation. The four-component Dirac spinor is composed of two two-component Weyl fermions of opposite chirality. Under magnetic field applied parallel to the electric field $\vec{E}\|\vec{B}$, charge is predicted to flow between the Weyl nodes resulting in negative magnetoresistance, a phenomenon known as the Adler-Bell-Jackiw anomaly~\cite{Xiong413}. Furthermore, the mirror anomaly~\cite{PhysRevLett.120.016603} was predicted in DSMs. Although the degeneracy of the Dirac point (DP) is not protected in topological sense due to its zero net Chern number, it can still be protected by the space group symmetries, e.g. $C_{4v}$~\cite{PhysRevB.88.125427}, $C_{6v}$~\cite{PhysRevB.85.195320} and non-symmorphic symmetries~\cite{PhysRevB.92.165120, RevModPhys.90.015001}, and therefore referred to as the {\it symmetry-protected} degeneracy. Several materials realizing such symmetry-protected DSM phase have been proposed theoretically~\cite{PhysRevLett.108.140405, PhysRevB.85.195320, PhysRevB.88.125427, Du2015, PhysRevMaterials.1.044201, Le2018} and confirmed experimentally~\cite{Liu864,PhysRevLett.108.140405, Neupane2014}. Notable examples of DSMs are Na$_3$Bi and Cd$_3$As$_2$ with DPs protected by $C_{6v}$ and $C_{4v}$ rotation symmetries, respectively. However, Na$_3$Bi oxidizes in air easily while arsenic is poisonous thus limiting applications of Cd$_3$As$_2$. Therefore, searching for new 3D DSMs that are stable at ambient conditions and are less toxic is of both fundamental and technological importance. 

In this work, we report an investigation of a new predicted Dirac semimetal MgTa$_2$N$_3$. By means of first-principles calculations and symmetry analysis we address the electronic structure and topological properties of this material. We further predicted some properties of the topological phase that can be measured experimentally and show how different topological phase transitions can be attained by  breaking crystalline symmetries.

\begin{figure*}
    \centering
    \includegraphics[width=16cm]{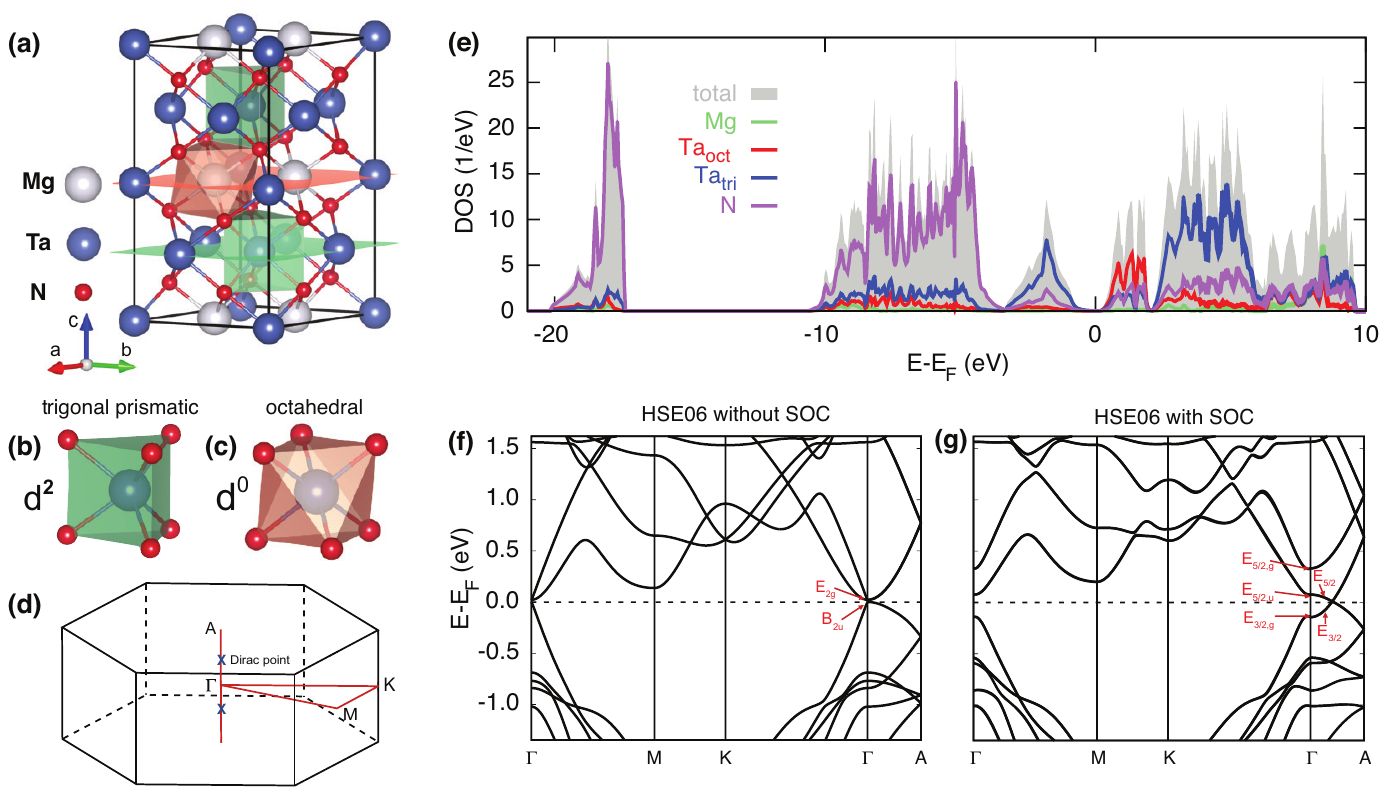}
    \caption{(a) Crystal structure  of MgTa$_{2}$N$_3$ with  (b) trigonal prismatic (c) octahedral coordination spheres of metal ions being indicated in green and orange, respectively.
  (d) The Brillouin zone (BZ) of MgTa$_{2}$N$_3$ showing the location of the Dirac points.  (e) Total and projected density of states  calculated using the HSE06 functional without SOC.
 (f,g) Band structures calculated using the HSE06 functional (f) without and (g) with SOC. }
    \label{fig1}
\end{figure*}

% crystal structure description
The synthesis of MgTa$_2$N$_3$ was reported by Brokamp and Jacobs in 1991~\cite{BROKAMP1992325}. The crystal structure of MgTa$_2$N$_3$ belongs to space group $P6_3/mcm$ with lattice constants $a=5.205$~\AA  \ and $c=10.425$~\AA. It consists of alternating layers of Ta atoms with trigonal prismatic and octahedral coordination spheres (below referred to as tri-Ta and oct-Ta, respectively), similar to numerous ABX$_2$ layered oxides~\cite{ABX2}, oxynitrides, and nitrides such as ScTaN$_2$~\cite{ScTaN2}. Two Mg atoms substitute two Ta atoms in the octahedral layer, suggesting the  (Ta$^{5+}_{\rm oct}$)$_2$(Mg$^{2+}$)$_4$(Ta$^{3+}_{\rm tri}$)$_6$N$^{3-}_{12}$ ionic picture with $d^0$ and $d^2$ electronic configurations of oct-Ta and tri-Ta, respectively (Figs.~\ref{fig1}(a--c)). 

% method description
Our first-principles band-structure calculations are performed within the density functional theory framework
using VASP (Vienna Ab initio Simulation Package)~\cite{PhysRevB.54.11169, PhysRevB.59.1758}. The approach relies on all-electron projector augmented wave (PAW) basis sets~\cite{PhysRevB.50.17953} combined with the generalized gradient approximation (GGA) with exchange-correlation functional of Perdew, Burke and Ernzerhof (PBE)~\cite{PhysRevLett.77.3865} and the Heyd–Scuseria–Ernzerhof (HSE06) hybrid functional~\cite{HSE06}.
Both PBE and HSE06 calculations predict the existence of Dirac point degeneracies in the band structures when spin-orbit coupling (SOC) is taken into account. 
The PBE functional calculations, however, show the presence of an additional Fermi surface pocket at the $M$ point that could be caused by the underestimated correlation effect. In the rest of the paper the Heyd–Scuseria–Ernzerhof (HSE06)~\cite{HSE06} hybrid functional will be used to take into account the non-local potential.  Detailed comparison of the results obtained using PBE and HSE06 functionals  is presented in the Appendix~\ref{appendix:pbehse}. The cutoff energy for the plane wave expansion was set to 500~eV and a $k$-point mesh of $8\times8\times6$ was used in the bulk calculations. The WannierTools code~\cite{wanniertools} was used to investigate the topological properties and calculate the Landau levels (LLs) based on the maximal localized Wannier functions tight-binding model~\cite{RevModPhys.84.1419} that was constructed by using the Wannier90 package~\cite{wannier90} with Ta $5d$ atomic orbitals as projectors. The surface state spectra are calculated using the iterative Green's function method~\cite{0305-4608-15-4-009, wanniertools}. 

% DOS description
The total and atom-projected density of states (DOS)  calculated without taking into consideration the spin-orbit coupling are shown in Fig.~\ref{fig1}(e). A direct band gap of 21~meV is seen at the $\Gamma$ point. The states of Mg are located entirely above the Fermi level showing the fully ionic character of Mg$^{2+}$ ions.  The $2s$ states of N atoms are located between $-$21~eV to $-$17~eV. The energy range between $-$11~eV and $-$4~eV is dominated by the $2p$ states of N atoms and a fraction of Ta $d$ states. Above the energy of $-$4~eV, the $d$ states of oct-Ta are located above the Fermi level and there is a peak originating from tri-Ta states between $-$4~eV to 0~eV, which is consistent with the suggested $d^0$ and $d^2$ electronic configurations of oct-Ta and tri-Ta ions. 

%band structure description
The band structure of MgTa$_{2}$N$_3$ calculated without taking SOC into account is shown in Fig.~\ref{fig1}(f). The valence band maximum (VBM) and the conduction band minimum (CBM) are located at the $\Gamma$ point. The VBM is composed of the $d_{z^2}$ orbitals of tri-Ta atoms and belongs to the $B_{2u}$ irreducible representation of the $D_{6h}$ point group \cite{altmann1994}. The CBM is mainly composed of the $d_{xy}$ and  $d_{x^2-y^2}$ orbitals of oct-Ta atoms and belongs to the $E_{2g}$ irreducible representation of $D_{6h}$ \cite{altmann1994}. The spin-orbit coupling splits the $E_{2g}$ representation into $E_{5/2,g}$ and $E_{3/2,g}$ representations resulting in the separation of the 4-fold degenerate bands into pairs of 2-fold degenerate bands, eventually leading to band inversion between the $E_{5/2,u}$ and $E_{3/2,g}$ bands (see Fig.~\ref{fig1}(g)). Along the $\Gamma-$A direction, two bands close to the Fermi level belong to the $E_{3/2}$ and $E_{5/2}$ irreducible representations of the little group of $C_{6v}$\cite{altmann1994}. According to the Schur's lemma the two bases belonging to two different irreducible representations are orthogonal to each other, hence the two bands cross without opening a gap resulting in a Dirac point band degeneracy. In MgTa$_{2}$N$_3$ the two DPs are located at (0, 0, $\pm$0.203$\frac{\pi}{c}$). 

\begin{figure}[b]
    \centering
    \includegraphics[width=8cm]{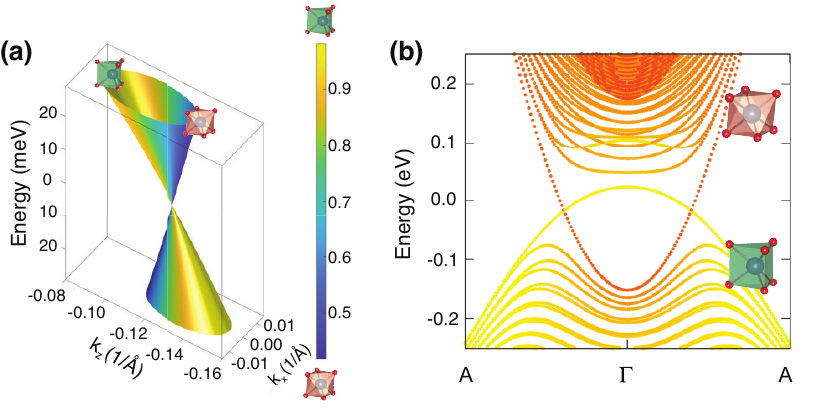}
    \caption{ (a) Dirac cone band dispersion showing the orbital weights of the tri-Ta and oct-Ta ions. (b) Landau levels calculated for magnetic field ${\vec B}$ applied parallel to the $c$ axis.}
    \label{fig2}
\end{figure} 

% subdeduction table
\begin{table*}[ht]
\caption{Subduction table for the little group of a $k$ point along the $\Gamma-A$ direction (2nd row) and the compatible relation of its irreducible representations $E_{3/2}$ and $E_{5/2}$ composing the Dirac cone states. The first row is the little group at the $\Gamma$ point which is the supergroup of group in the second row. $E_{1/2}$, $E_{3/2}$ and $E_{5/2}$ are two-dimensional representations, while all others are one-dimensional representations. Red color indicates the entries containing inversion symmetry. }\label{table:deduction}
\begin{center}
 \begin{adjustbox}{max width=1.0\textwidth, max totalheight=\textheight,keepaspectratio}
\begin{tabular}{|*{7}{c|}}
\hline
$\Gamma$ & ${\bf \color{red} D_{6h}}$, $C_{6v}$ & ${ \bf\color{red}{D_{3d}}}$, $D_{3h}$, $C_{3v}$ & $D_6$, $C_6$ &  ${\bf\color{red}S_6}$, $D_3$, $C_3$ & ${\bf\color{red}D_{2h}}$, $C_{2v}$ &$\bf\color{red}C_{2h}$, $C_s$, $C_2$\\
\hhline{*{7}-}
$\Gamma$-A&$C_{6v}$  & $C_{3v}$  &$C_{6}$ & $C_{3}$ & ${C_{2v}}$& $C_s (C_2)$\\
\hhline{*{7}-}
\multirow{2}{*}{irrep.}&$E_{3/2}$ & $ {}^1E_{3/2} \oplus {}^2E_{3/2}$ &$ {}^1E_{3/2} \oplus {}^2E_{3/2}$ & ${}^2A_{3/2}$ & $E_{1/2}$ &  $ {}^1E_{1/2} \oplus {}^2E_{1/2}$  \\
&$E_{5/2}$ & $E_{1/2}$ & $ {}^1E_{5/2} \oplus {}^2E_{5/2}$  &  $ {}^1E_{1/2} \oplus {}^2E_{1/2}$ & $E_{1/2}$ &  $ {}^1E_{1/2} \oplus {}^2E_{1/2}$  \\
\hhline{*{7}-}
Phase&Dirac & {\bf\color{red} Dirac}/Triple & Weyl & {\bf\color{red} Dirac}/Weyl & TI & TI or Weyl\\
\hline
\end{tabular}
\end{adjustbox}
\end{center}
\label{default}
\end{table*}

From the DOS and symmetry analysis it follows that the two energy bands forming the Dirac cones, referred to as the $E_{3/2}$ and $E_{5/2}$ bands hereafter, are composed of atomic orbitals of either oct-Ta or tri-Ta atoms belonging to the two distinct layers in the crystal structure. The weights of the oct-Ta and tri-Ta $d$ orbitals in the Dirac cone band in the $k_x-k_z$ plane are shown in Fig.~\ref{fig2}(a). One can also see that the Dirac cone is anisotropic in the $k_x-k_z$ plane according to the difference of the lattice constants $a$ and $c$.  {The calculated Fermi velocities of the Dirac fermion charge carriers  \cite{SM} are $v_x=v_y=2.8$~eV$\cdot$\AA, $v_z({E_{5/2}})$=0.95~eV$\cdot$\AA, $v_z({E_{3/2}})$=2.38~eV$\cdot$\AA. These Fremi velocities are comparable to that in Na$_3$Bi \cite{Liu864, PhysRevB.94.085121,Kushwaha2015}, but smaller than in Cd$_3$As$_2$ \cite{Liu2014,Liang2014,Neupane2014}}. In the $k_x$ direction, the weights are symmetrical as $\rho_n(k_x, k_y, k_z)= \rho_n(-k_x, k_y, k_z)$, where $\rho_n({\bf k}) = \sum_{\alpha} \langle \psi_{n {\bf k}}|\varphi_{\alpha}\rangle\langle\varphi_{\alpha}|\psi_{n{\bf k}}\rangle$, $\psi_{n {\bf k}}$ is the Bloch wave function and $\varphi_{\alpha}$ is the $d$ atomic orbital of tri-Ta or oct-Ta atoms. 
Along the $k_z$ direction, the weight is larger for tri-Ta  $d$ orbitals close to the $\Gamma$ point in the upper half of the conical intersection, while the opposite is true in the lower cone. It is worth mentioning that larger weights of tri-Ta orbitals are due to the fact there three times as many tri-Ta atoms than the oct-Ta atoms in the crystal structure of MgTa$_{2}$N$_3$.  

Figure~\ref{fig2}(b) shows the weighted Landau levels (LLs) formed from the energy bands upon applying magnetic field along the $c$ axis. It is known that the two zeroth LLs could lead to the chiral anomaly. The upward and downward parabolic curves are mostly associated with the atomic orbitals of oct-Ta and tri-Ta, respectively. The LLs provide a way of studying the layer-resolved features. 

% study the topological phase transition due to the symmetry protection.
Since the crystal symmetries protect the Dirac points in MgTa$_{2}$N$_3$, it is important to address possible topological phase transitions realized upon breaking these symmetries. Here, we consider breaking only the spatial symmetries while preserving the time-reversal symmetry. One qualitative approach to this question consists in studying the compatibility relationships of the two irreducible representations $E_{3/2}$ and $E_{5/2}$ when deducing the little group $C_{6v}$ of the DP. The corresponding subduction table~\cite{altmann1994} is shown in Table~\ref{table:deduction}. From this table, we can divide the possible ways of symmetry breaking into two groups, with and without inversion symmetry. In presence of inversion symmetry all energy bands are doubly degenerate according to the Kramers theorem, hence only the DSM or topological insulator (TI) phases can be realized. Table \ref{table:deduction} shows that $C_3$ and inversion symmetries are sufficient to protect the DP degeneracy. Breaking the $C_3$ symmetry results in the transition of DSM into the TI phase. Without inversion symmetry, the double degeneracy can be lifted by breaking the $C_3$ and mirror symmetries. We distinguish three types of symmetry breaking referred to as types  A, B and C.  Type-A symmetry breaking preserves the $C_3$ symmetry while breaking the vertical mirror $\sigma_v$ symmetry or the $\sigma_d$ symmetry, resulting in the $C_{3v}$ group. The two-dimensional representation $E_{3/2}$ splits into two one-dimensional representations $ {}^1E_{3/2}$ and  ${}^2E_{3/2}$, while $E_{5/2}$ changes into the two-dimensional representation $E_{1/2}$. Eventually, the DPs splits into two triple nodal points (TNPs)~\cite{PhysRevLett.117.076403,PhysRevX.6.031003, PhysRevB.93.241202, PhysRevB.94.165201, PhysRevLett.119.256402, Chang2017, Lv2017}. Type-B symmetry breaking preserves the $C_3$ symmetry while breaking all mirror symmetries $\sigma_v$ and $\sigma_d$, thus resulting in the $C_6$ group or the $C_3$ group upon further breaking the $C_2$ symmetry.
The two two-dimensional representations $E_{3/2}$ and $E_{5/2}$ split into four different one-dimensional representations resulting in the separation of DPs into four symmetry-protected Weyl points. Type-C symmetry breaking eliminates the $C_3$ symmetry resulting in the $C_{2v}$ group or the $C_s(C_2)$ group upon further breaking the $C_2$($\sigma_v$) symmetry. Here, the $E_{3/2}$ and $E_{5/2}$ representations split into the same representations, 
leading to the strong TI phase.  It is worth mentioning that $C_3$ symmetry is not sufficient for protecting a DP, and the presence of inversion or six vertical mirror symmetries is required. This is at odds with the conclusion of Ref.~\onlinecite{PhysRevB.85.195320} where it is claimed that the $C_3$ symmetry is sufficient to protect the DPs. 

\begin{figure*}[t]
    \centering
    \includegraphics[width=16cm]{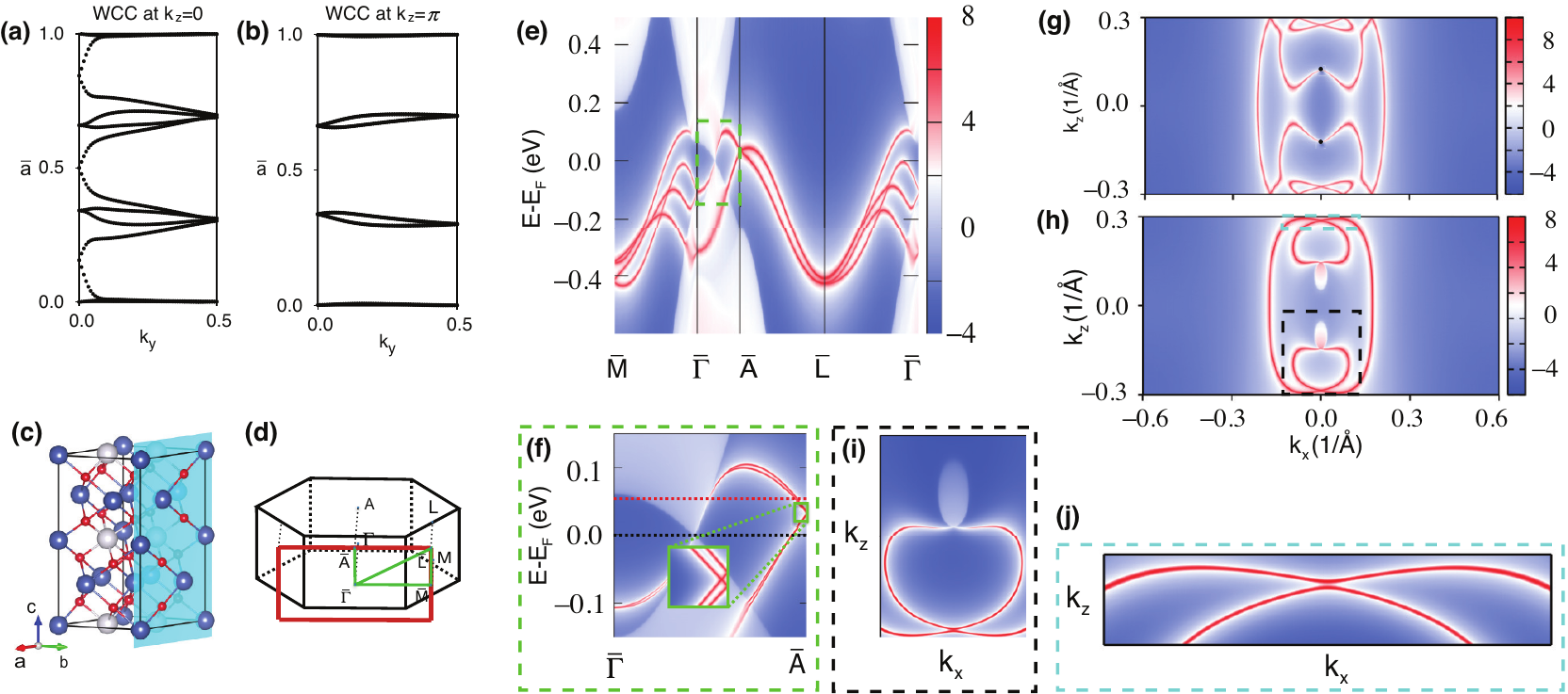}
    \caption{(a),(b) Evolution of hybrid Wannier charge centers at $k_z=0$ and $k_z=\pi$, respectively. (c) Crystal structure of MgTa$_2$N$_3$ showing the (010) surface plane. (d) The corresponding two-dimensional surface BZ (red rectangle). (e) (100) surface states along the $k$-point path shown as green lines in panel (d). Details of the surface states in a narrow range of $k$ and $E$ (green region is zoomed in panel (f)).  (g),(h) Surface states at energies $E-E_F = 0.0$~eV and $E-E_F=0.05$~eV, respectively, with the corresponding details shown in panels (i) (black box) and (j) (cyan box). }
    \label{fig3}
\end{figure*} 

In order to get a quantitative insight, we build a universal $4\times4$ $k\cdot p$ model for the point groups in the first row of Table \ref{table:deduction}. Here, we only list matrix elements up to quadratic term in the diagonal matrix elements and up to  linear term of $k_z$ in the off-diagonal matrix elements
\begin{align}
\begin{adjustbox}{max width=0.48\textwidth}
$H({\bf k})= \epsilon_0({\bf k}) +
\left(
\begin{array}{cccc}
M({\bf k})+B_1(k_z) & B_2(k_z)  & D_1(k_z) & A_1(k_+,k_-)\\
B^*_2(k_z) & M({\bf k})-B_1(k_z) &  A_2(k_+,k_-) & D_2(k_z) \\
 D_1^*(k_z) &  A^*_2(k_+,k_-)  & -M({\bf k})+B_1'(k_z) & B_2'(k_z) \\
A^*_1(k_+,k_-) & D^*_2(k_z) &  B'^{*}_2 (k_z) & -M({\bf k})-B_1'(k_z)  \\
\end{array}
\right)$,
\end{adjustbox}
\end{align}
where $\epsilon_0({\bf k})=C_0+C_1k_z^2+C_2(k_x^2+k_y^2)$, $k_{\pm}=k_x\pm ik_y$, $A_1$, $A_2$ are the linear combinations of $k_+$ and $k_-$, and $M({\bf k})=M_0-M_1k_z^2-M_2(k_x^2+k_y^2)$. $M_0\cdot M_1>0$ is the condition of band inversion along the $\Gamma-A$ direction.  The details of construction can be found in the Appendix \ref{appendix:kp}.% . 
Under the $D_{6h}$ or $C_{6v}$ groups that preserve the $C_3$, $\sigma_v$ and $\sigma_d$ symmetries, all linear terms of $k_z$ vanish as $B_1(k_z)=B_2(k_z)=B'_1(k_z)=B'_2(k_z)=D_1(k_z)=D_2(k_z)=0$, resulting in two Dirac points at $k_{DPs}=(0,0,\pm\sqrt{\frac{M_0}{M_1})}$. For the type-A symmetry breaking, only $B_1(k_z)= Bk_z$ or $B_2(k_z)=Bk_z$ are present, resulting in two pairs of TNPs at $k_{TNPs}=\pm(0,0,\frac{B\pm\sqrt{B^2-16M_0M_1}}{4M_1})$. For the type-B symmetry breaking, only $B_1(k_z)=Bk_z$ and $B_1'(k_z)=B'k_z$ are present. There are four pairs of WPs located at $k_{WPs}=\pm(0,0,\frac{\pm B_0\pm\sqrt{B^2_0+16M_0M_1}}{4M_1})$ with $B_0=B'\pm B$. For the type-C symmetry breaking, the coupling terms between the two bands forming the Dirac cone are $D_1(k_z)=D_2(k_z)=Dk_z$, which leads to a finite effective mass and results in the Hamiltonian characteristic of a strong TI phase. 

% topological properties and surface states
We now address the topological properties of MgTa$_2$N$_3$. The evolution of hybrid Wannier charge centers (WCCs) ~\cite{PhysRevB.84.075119, PhysRevB.83.035108, wanniertools} in the time-reversal invariant planes $k_z=0$ and $k_z=\pi/c$ is shown in Figs.~\ref{fig3}(a) and \ref{fig3}(b), respectively. The corresponding $\mathcal{Z}_2$ numbers are 1 and 0, indicating that band inversion between $E_{3/2}$ and $E_{5/2}$ bands along the $\Gamma-A$ direction is topologically non-trivial. The non-trivial $\mathcal{Z}_2$ number at $k_z=0$ plane is expected to result in one surface Dirac cone at the $\bar{\Gamma}$ point on the side surfaces, e.g. the (010) surface shown in Fig.~\ref{fig3}(c).  The space group of the (010) surface is $Pma2$ (No. 28), whose generators are the two-fold rotation $C_{2y}$ with axis along the $y$ direction and the $G_x=\{\sigma_x|(0,0,1/2)\}$ glide symmetry. The latter symmetry leads to the hourglass surface states (SSs)~\cite{PhysRevLett.115.126803,Bzdusek2016,Wang2016,Wang2017} along the $\bar{\Gamma}-\bar{A}$ direction. The calculated momentum-resolved surface density of states along the high-symmetry line in the 2D BZ is shown in Fig.~\ref{fig3}(e). The hourglass dispersion of the SSs is shown in the inset of Fig.~\ref{fig3}(f), while Fig.~\ref{fig3}(g) shows the iso-energy plot of the SS spectrum at the Fermi energy $E-E_F=0$. One can observe the following two features: first, the presence of a long Fermi arc linking the two DPs; and second, the double degeneracy of bands at the 2D BZ boundary due to the glide symmetry $G_x$.

The DPs discussed here could be considered as the combination of two WPs with opposite chirality. According to the linking rules the Fermi arc surface state should connect  WPs of different chirality, hence in a DSM the Fermi arc could originate and terminate at the same DP or link together two different DPs. However, such Fermi arcs are not topologically protected as discussed in Refs.~\onlinecite{Kargarian8648,Le2018, PhysRevB.97.165129}. The degeneracy can be lifted along the $\bar{\Gamma}-\bar{A}$ direction if the corresponding term is present, and our symmetry analysis (see Appendix \ref{appendix:kp}) shows that such term naturally exists for the side surface of MgTa$_2$N$_3$. The observed splitting is shown in Fig.~\ref{fig3}(f). 
In order to observe the ``unprotected'' Fermi arc states we plot the iso-energy surface-state spectrum at $E-E_F=0.05$ (Fig.~\ref{fig3}~(h)). Fig.~\ref{fig3}(i) provides the details of the surface states that reveal a ``candlelight'' shaped Fermi arcs that originate and terminate at the same Fermi pocket. The lack of connection between the ``candlelight'' Fermi arcs and other states is shown in Fig.~\ref{fig3}(j), which covers the momentum range indicated by the cyan box in Fig.~\ref{fig3}(h).

In summary, based on first-principles calculations and effective model analysis we predicted an ideal Dirac semimetal phase in MgTa$_2$Ta$_3$, a material that has been synthesized previously. The Dirac cone band degeneracies are composed of two atomic $d$ orbitals originating from two distinct layers of Ta atoms with different coordination. We then analyzed possible topological phase transitions that can be realized by breaking lattice symmetries. It was found that the Dirac semimetal phase can be transformed into a number of distinct topological phases, namely the triple nodal point metal, Weyl semimetal, strong topological insulator, without breaking the time-reversal symmetry. In practice, such symmetry breaking can be realized by strain along different directions or alloying. The topologically protected surface states as well as the Fermi arcs lacking such protection were studied in detail. We point out that the ``candlelight'' shaped unprotected Fermi arcs can be detected in ARPES measurements. 

We thank X. Dai, H. M Weng and V. M Katukuri for helpful discussions.  Q.W., O.V.Y. acknowledge support by NCCR Marvel. Q.W and C.P acknowledge support by Matthias Troyer. First-principles calculations were performed at the Swiss National Supercomputing Centre (CSCS) under project s832 and the facilities of Scientific IT and Application Support Center of EPFL. Z.S was supported  by National Natural Science Foundation of China, the National  973 program of China (Grant No. 2013CB921700) 

{\it Note:} Another work addressing the same material~\cite{PhysRevLett.120.136403} appeared when the present manuscript was in preparation. Our work was presented at the APS March meeting 2018~\cite{aps2018wu} prior to the publication of Ref.~\onlinecite{PhysRevLett.120.136403}.

\appendix

%section PBE VS HSE
\section{\label{appendix:pbehse} Comparison of the PBE and HSE06 band structures}

\begin{figure*}[htp]
    \centering
    \includegraphics[width=10.5cm]{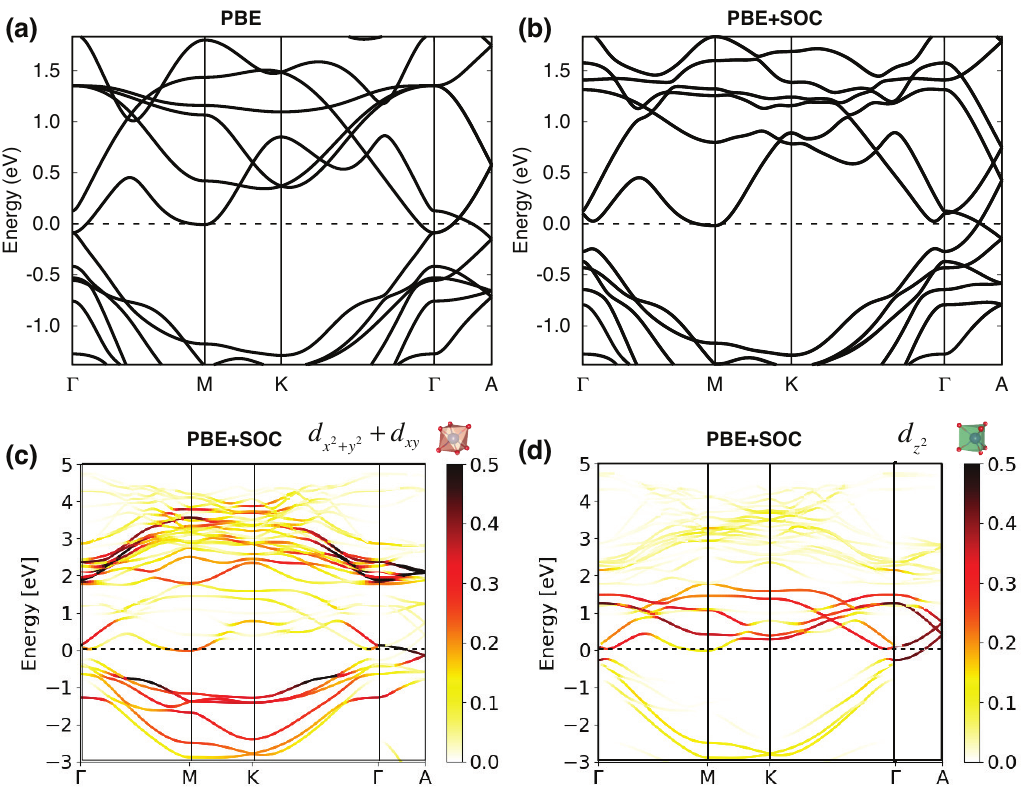}
    \caption{(a),(b) The calculated band structures of MgTa$_2$N$_3$ within the PBE function calculations (a) without and (b) with SOC. (c) Projected band structure with the weights of oct-Ta $d_{x^2+y^2}+d_{xy}$ orbitals indicated. (d) Projected band structure with the weights of tri-Ta $d_{z^2}$ orbitals indicated.  \label{pbe}}
    \label{fig3}
\end{figure*}

In the main text, we discuss the results obtained using the HSE06 hybrid functional. Here, we provide the band structures obtained using the PBE functional for comparison (see Fig.~\ref{pbe}). At the $M$ point, there is an additional hole pocket which is absent in the hybrid functional calculations.  The PBE band structure without SOC is metallic in contrast to the gapped band structure in  the HSE06 calculations. The open-source code {\it PyProcar}~\cite{pyprocar} is used to generate the weighted band structures shown in Figs.~\ref{pbe}(c) and \ref{pbe}(d).

%section appendix:kp
\section{\label{appendix:kp} $k\cdot p$ models for the bulk material at different symmetry-breaking ditortions} 

In this section, we construct several $k\cdot p$ models at the $\Gamma$ point for different point groups listed in Table~\ref{table:deductiongamma}. The spatial and time-reversal symmetries are taken into consideration. All $k\cdot p$ models are constructed using the  {\it kdotp-symmetry} open source package.~\cite{Gresch2017} Details on the procedure for building a $k\cdot p$ model are given in Refs.~\onlinecite{1367-2630-19-3-035001, Feng2017}. Symmetrized basis sets (see p.~74 of Ref.~\onlinecite{altmann1994point}) are chosen according to the irreducible representations in Table~\ref{table:deductiongamma}. Here, we list the terms linear and quadratic in $k_x, k_y, k_z$ as well as cubic terms involving the combination of $k_z$ with $k_x^2, k_y^2$. For the sake of compact notations, we divide $k\cdot p$ models into the diagonal part $H_0({\bf k})$ that is the same for all  point groups and $H'(\bf{k})$ such that
\begin{widetext}
\begin{align}
H({\bf{k}})=H_0({\bf k})+H'({\bf k})
\end{align}
with $H_0({\bf k})$
\begin{align}
\begin{adjustbox}{max width=\textwidth}
$H_0({\bf k})= \left(
\begin{array}{cccc}
\epsilon_0({\bf k})+M({\bf k}) & 0 & 0 & 0\\
0 & \epsilon_0({\bf k})-M({\bf k})&  0 & 0 \\
 0 &  0 & \epsilon_0({\bf k})+M({\bf k}) & 0 \\
0 & 0 &0& \epsilon_0({\bf k})-M({\bf k}) \\ 
\end{array}
\right) ,$
\end{adjustbox}
\end{align}
\end{widetext}
where $\epsilon_0({\bf k})=C_0+C_1k_z^2+C_2(k_x^2+k_y^2)$ and $M({\bf k})=M_0-M_1k_z^2-M_2(k_x^2+k_y^2)$.

\begin{table*}[!htpb]
\caption{Subduction table for the little group at the $\Gamma$ point and the compatible relation of its irreducible representations $E_{3/2,g}$ and $E_{5/2,u}$ composing the Dirac cone. 
This table is the simplified version of Table 35.9 in Ref.~\onlinecite{altmann1994point}.}\label{table:deductiongamma}
\begin{center}
 \begin{adjustbox}{max width=\textwidth, max totalheight=\textheight,keepaspectratio}
\begin{tabular}{cccccccc}
\hline\hline
$D_{6h}$ & $D_{3h}$ & $D_{3d}$  & $D_6$ & $D_3$ & $D_{2h}$ & $C_{2v}$ & $C_{6h}$  \\
\hline
$E_{3/2,g}$ &$E_{3/2}$ &${}^1E_{3/2,g}\oplus{}^2E_{3/2,g} $&$E_{3/2}$ &${}^1E_{3/2}\oplus{}^2E_{3/2} $ & $E_{1/2,g}$  & $E_{1/2}$ & ${}^1E_{3/2,g}\oplus{}^2E_{3/2,g} $\\ 
$E_{3/2,u}$ &  $E_{1/2}$&$E_{1/2,u}$ & $E_{5/2}$ & $E_{1/2}$ & $E_{1/2,u}$ & $E_{1/2}$ & ${}^1E_{5/2,u}\oplus{}^2E_{5/2,u} $\\
\hline 
$C_{6v}$  & $C_{3v}$  &$C_{6}$ & $C_{3}$ & $C_{2v}$ & $C_s (C_2)$ & $S_6$ \\
\hline
$E_{3/2}$ & $ {}^1E_{3/2} \oplus {}^2E_{3/2}$ &$ {}^1E_{3/2} \oplus {}^2E_{3/2}$ & ${}^2A_{3/2}$ & $E_{1/2}$ &  $ {}^1E_{1/2} \oplus {}^2E_{1/2}$ & $2A_{3/2,g}$ \\
$E_{5/2}$ & $E_{1/2}$ & $ {}^1E_{5/2} \oplus {}^2E_{5/2}$  &  $ {}^1E_{1/2} \oplus {}^2E_{1/2}$ & $E_{1/2}$ &  $ {}^1E_{1/2} \oplus {}^2E_{1/2}$  & ${}^1E_{1/2,u}\oplus{}^2E_{1/2,u} $\\
\hline\hline
\end{tabular}
\end{adjustbox}
\end{center}
\label{default}
\end{table*}% 

{$\boldsymbol{D_{6h}}$}. The little group of MgTa$_2$N$_3$ for the $\Gamma$ point is $D_{6h}$, hence the constructed $k\cdot p$ model is 
\begin{widetext}
\begin{align}
\begin{adjustbox}{max width=\textwidth}
$H'_{D_{6h}}({\bf k})=
\left(
\begin{array}{cccc}
0 & 0 &  iF k_z k_+^2 & iA k_+ \\
0 & 0&  -iA k_-  & iF k_z k_-^2 \\
-iF k_z k_-^2 &  iA k_+ & 0& 0 \\
-iA k_- &    -iF k_z k_+^2  & 0 & 0 \\ 
\end{array}
\right) ,$
\end{adjustbox}
\end{align}
\end{widetext}
where $k_{\pm}=k_x\pm ik_y$, $i$ is the imaginary unit. This $k\cdot p$ model is the same as that of Na$_3$Bi at the $\Gamma$ point up to a unitary transformation. The Dirac points occur only when $M_0M_1>0$, which is the condition for band inversion. The parameters obtained by fitting the energy bands to the result of first-principles calculations are $C_0=-0.0336$~eV, $C_1=2.9058$~eV~\AA$^2$,$C_2=-3.8226$~eV~\AA$^2$, $M_0=-0.1098$~eV, $M_1=-7.3012$~eV~\AA$^2$, $M_2=-15.5233$~eV~\AA$^2$, A=$2.9129$~eV~\AA.

{$\boldsymbol{D_{3h}}$}. By breaking the inversion symmetry $i$, the mirror symmetries $\sigma_{d1}$,  $\sigma_{d2}$,  $\sigma_{d3}$ and the $C_6$ rotation symmetry, $D_{6h}$ deduces to the $D_{3h}$ group. The $k\cdot p$ model is therefore 

\begin{widetext}
\begin{align*}
\begin{adjustbox}{max width=\textwidth}
$H'_{D_{3h}}({\bf k})=
\left(
\begin{array}{cccc}
0 & iBk_z+iGk_zk^2_{\parallel} & D_2 k_zk_-+i F_1k_z k^2_+ & iA k_++D_1k^2_- \\
-iB k_z -iGk_zk^2_{\parallel}  & 0&  -iA k_--D_1k^2_+  & D_2 k_zk_++i F_1k_z k^2_- \\
D_2 k_z k_++i F_1k_z k^2_-  &  iA k_+-D_1k^2_-  & 0& i F_2 k_zk^2_- \\
-iA k_-+D_1k^2_+  &    D_2 k_z k_- -i F_1k_z k^2_+ &  -iF_2 k_zk^2_+ & 0 
\end{array}
\right) .$
\end{adjustbox}
\end{align*}
\end{widetext}

{$\boldsymbol{D_{3d}}$}. By breaking the horizontal mirror symmetry $\sigma_h$, the vertical mirror symmetries $\sigma_{v1}$,  $\sigma_{v2}$,  $\sigma_{v3}$ and the $C_6$ rotation symmetry, $D_{6h}$ deduces to the $D_{3d}$ group. It is worth mentioning that the inversion symmetry is preserved. The corresponding $k\cdot p$ model is 
\begin{widetext}
\begin{align*}
\begin{adjustbox}{max width=\textwidth}
$H'_{D_{3d}}({\bf k})=
\left(
\begin{array}{cccc}
0 & 0 & A_1k_-+F_1k_zk^2_+ & -A_1k_+ +F_1k_zk^2_- \\
0 & 0 & A_2 k_-  +F_2k_zk^2_+ &  A_2 k_+ -F_2k_zk^2_- \\
A_1k_++F_1k_zk^2_-  & A_2 k_+  +F_2k_zk^2_- & 0 & 0 \\
-A_1k_- +F_1k_zk^2_+ &  A_2 k_- -F_2k_zk^2_+ & 0 & 0 
\end{array}
\right) .$
\end{adjustbox}
\end{align*}
\end{widetext}
In this $k\cdot p$ model the energy bands along $\Gamma-A$ are double degenerate and the Dirac cone is not split. Although $C_{3v}$ is the maximum subgroup of $D_{3d}$ and $D_{3h}$, there are certain differences. The difference is the $PT$ symmetry which is the combination of inversion and time-reversal symmetries. The little group of the $k$ point along $\Gamma-A$ is $C_{3v}$ plus the $PT$ symmetry when the little group of the $\Gamma$ point is $D_{3d}$. All the bands become double degenerate. Hence, breaking the inversion symmetry and the vertical mirror symmetries $\sigma_{v1}$, $\sigma_{v2}$,  $\sigma_{v3}$ (or  $\sigma_{d1}$  $\sigma_{d2}$,  $\sigma_{d3}$) while keeping other vertical mirror symmetries $\sigma_{d1}$  $\sigma_{d2}$,  $\sigma_{d3}$ (or  $\sigma_{v1}$, $\sigma_{v2}$,  $\sigma_{v3}$ ) are the keys for the reaching the triple nodal point phases.

{$\boldsymbol{C_{6h}}$}.  This group can be obtained from $D_{6h}$ by breaking all vertical mirror symmetries and the two-fold rotation symmetries in the $xy$ plane. The inversion symmetry is preserved. The $k\cdot p$ model is 
\begin{widetext}
\begin{align*}
\begin{adjustbox}{max width=\textwidth}
$H'_{C_{6h}}({\bf k})=
\left(
\begin{array}{cccc}
0 & 0 & A_1k_-+F_1k_zk^2_+ & -A_1k_+ +F_1k_zk^2_- \\
0 & 0 & A_2 k_-  +F_2k_zk^2_+ &  A_2 k_+ -F_2k_zk^2_- \\
A_1k_++F_1k_zk^2_-  & A_2 k_+  +F_2k_zk^2_- & 0 & 0 \\
-A_1k_- +F_1k_zk^2_+ &  A_2 k_- -F_2k_zk^2_+ & 0 & 0 
\end{array}
\right) .$
\end{adjustbox}
\end{align*}
\end{widetext}
There is no term that can break the degeneracy of the Dirac point. This confirms that breaking all vertical mirror symmetries wouldn't lift the degeneracy of the Dirac point in the presence of both inversion symmetry and the $C_3$ symmetry. 

%S6
{$\boldsymbol{S_{6}}$}. In $D_{6h}$, {${S_{6}}$} is the smallest subgroup that has both inversion symmetry and the $C_3$ symmetry but lacks mirror symmetry. The $k\cdot p$ model is 
\begin{widetext}
\begin{align}
\begin{adjustbox}{max width=\textwidth}
$H'_{S_{6}}({\bf k})=
\left(
\begin{array}{cccc}
0 & 0 & A_1k_-+F_1k_zk^2_- & A_2k_- +F_2k_zk^2_+ \\
0 & 0 & A_2 k_+  +F_2k_zk^2_- &  -A_1 k_- -F_1k_zk^2_+ \\
A_1k_- +F_1k_zk^2_+  & A_2 k_-  +F_2k_zk^2_+ & 0 & 0 \\
A_2k_+ +F_2k_zk^2_- &  -A_1 k_+ -F_1k_zk^2_- & 0 & 0 
\end{array}
\right) .$
\end{adjustbox}
\end{align}
\end{widetext}
The difference between $S_{6}$ and $C_{3h}$ is the presence of horizontal mirror symmetry $\sigma_h$ and the $C_{2z}$ rotation symmetry in the latter.  In this $k\cdot p$ model, the presence of inversion symmetry and the $C_3$ symmetry is sufficient for protecting the Dirac point. 

%C3
{$\boldsymbol{C_{3}}$}. {${C_{3}}$} is the maximum subgroup of $S_3$ after breaking inversion symmetry but keeping the $C_3$ symmetry. The $k\cdot p$ model is 
\begin{widetext}
\begin{align}
\begin{adjustbox}{max width=\textwidth}
$H'_{C_3}({\bf k})=
\left(
\begin{array}{cccc}
B_1k_z & B_2 k_z & (A_1+D_1k_z)k_-+F_1k_zk^2_+ & (A_2+D_2k_z)k_+ +F_2k_zk^2_- \\
B_2 k_z & -B_1k_z & (A_2-D_2k_z)k_-  -F_2k_zk^2_+ &  (-A_1+D_1k_z) k_+ +F_1k_zk^2_- \\
(A_1+D_1k_z)k_+ +F_1k_zk^2_-  &(A_2-D_2k_z) k_+  -F_2k_zk^2_+ & B'k_z & A_3 k_-\\
(A_2+D_2k_z)k_- +F_2k_zk^2_+ &  (-A_1+D_1k_z) k_- +F_1k_zk^2_+ &  A_3 k_+ &  -B'k_z 
\end{array}
\right) .$
\end{adjustbox}
\end{align}
\end{widetext}
The $B_1k_z$, $B_2k_z$ and $B'k_z$ terms lift the degeneracy of the two-fold degenerated bands. However, there are no hybridization linear terms involving $k_z$ between the two blocks. Therefore, the Dirac points will split into four pairs of Weyl points.

%C6
{$\boldsymbol{C_{6}}$}. {${C_{6}}$} is the supergroup of $C_3$ obtained by adding the two-fold rotation symmetry $C_{2z}$. The $k\cdot p$ model is 
\begin{widetext}
\begin{align}
\begin{adjustbox}{max width=\textwidth}
$H'_{C_6}({\bf k})=
\left(
\begin{array}{cccc}
B_1k_z & 0 & 0 & (A_2+D_2k_z)k_+ \\
0 & -B_1k_z & (A_2-D_2k_z)k_-  &  0\\
0& (A_2-D_2k_z) k_+  & B'k_z & A_3 k_-\\
(A_2+D_2k_z)k_- &  0 &  A_3 k_+ &  -B'k_z 
\end{array}
\right) .$
\end{adjustbox}
\end{align}
\end{widetext}
The $B_1k_z$ and $B'k_z$ terms lift the degeneracy of the two-fold degenerated bands. However, there are no hybridization linear terms involving $k_z$ between the two blocks.   Therefore, the Dirac points will split into four pairs of Weyl points as in the case of $C_3$ group. 

%D3
{$\boldsymbol{D_{3}}$}. {${D_{3}}$} is the supergroup of $C_3$ obtained by adding the two-fold rotation symmetry $C_{2x}$. The $k\cdot p$ model is 
\begin{widetext}
\begin{align}
\begin{adjustbox}{max width=\textwidth}
$H'_{D_3}({\bf k})=
\left(
\begin{array}{cccc}
0 & B_2k_z & A_1k_-+F_1k_zk^2_+ & -A_1k_+ F_1k_zk^2_- \\
B_2k_z & 0 & A_2 k_-  +F_2k_zk^2_+ &  A_2 k_+ -F_2k_zk^2_- \\
A_1k_++F_1k_zk^2_-  & A_2 k_+  +F_2k_zk^2_- & B'k_z & 0 \\
-A_1k_- +F_1k_zk^2_+ &  A_2 k_- -F_2k_zk^2_+ & 0 & B'k_z 
\end{array}
\right) .$
\end{adjustbox}
\end{align}
\end{widetext}
The $B_2k_z$ and $B'k_z$ terms lift the degeneracy of the two-fold degenerated bands. There are no hybridization linear terms involving $k_z$ between the two blocks. Hence, the Dirac points will split into four pairs of Weyl points as in the cases of $C_3$ and $C_6$ groups. 

%D6
{$\boldsymbol{D_{6}}$}. {${D_{6}}$} is the supergroup of $D_3$ obtained by adding adding the two-fold rotation symmetry $C_{2z}$. The $k\cdot p$ model is 
\begin{widetext}
\begin{align}
\begin{adjustbox}{max width=\textwidth}
$H'_{D_6}({\bf k})=
\left(
\begin{array}{cccc}
B_1 k_z & 0 &  iF_1k_zk^2_+  & i(A_1+D_1k_z)k_+\\
0 & -B_1 k_z & -i(A_1-D_1k_z) k_-  &   iF_1k_zk^2_- \\
  -iF_2k_zk^2_-  & i(A_1-D_1k_z) k_+ & B'k_z & A_3k_- \\
-i(A_1+D_1k_z)k_-  &  -iF_1k_zk^2_+ & A_3k_+ & B'k_z 
\end{array}
\right) .$
\end{adjustbox}
\end{align}
\end{widetext}
The $B_1k_z$ and $B'k_z$ terms lift the degeneracy of the two-fold degenerated bands,  but there are no hybridization linear terms involving $k_z$ between the two blocks. The Dirac points will split into four pairs of Weyl points as in the cases of $C_3$, $C_6$ and $D_3$ groups. 

%C6v
{$\boldsymbol{C_{6v}}$}. $C_{6v}$ is the subgroup of $D_{6h}$ breaking the inversion symmetry, all two-fold rotational symmetries in the $xy$ plane and the $\sigma_h$ horizontal symmetry. Vertical mirror symmetries and the $C_{6z}$ rotational symmetry are preserved. The corresponding $k\cdot p$ model is 

\begin{widetext}
\begin{align}
\begin{adjustbox}{max width=\textwidth}
$H'_{D_6}({\bf k})=
\left(
\begin{array}{cccc}
0 & 0 &  F_1k_zk^2_+  & (A_1+iD_1k_z)k_+\\
0 & 0 &(A_1+iD_1k_z) k_-  &   -F_1k_zk^2_- \\
  F_2k_zk^2_-  & (A_1-iD_1k_z) k_+ & 0 & iA_3k_- \\
(A_1-iD_1k_z)k_-  &  -F_1k_zk^2_+ & -iA_3k_+ & 0
\end{array}
\right) .$
\end{adjustbox}
\end{align}
\end{widetext}
Like the $k\cdot p$ models of $D_{6h}$, $D_{3d}$, $C_{6h}$ and $S_6$ groups, the Dirac points are retained in $C_{6v}$ even without the inversion symmetry due to the presence of 6 vertical mirror symmetries $\sigma_{v1}$,  $\sigma_{v2}$,  $\sigma_{v3}$, $\sigma_{d1}$,  $\sigma_{d2}$ and  $\sigma_{d3}$. 
 
%C3v
{$\boldsymbol{C_{3v}}$}. $C_{3v}$ is the subgroup of $C_{6v}$ obtained by breaking one of the vertical mirror symmetries.  The $C_{3z}$ rotational symmetry is preserved. The corresponding $k\cdot p$ model is 
\begin{widetext}
\begin{align}
\begin{adjustbox}{max width=\textwidth}
$H'_{C_3}({\bf k})=
\left(
\begin{array}{cccc}
B_1k_z  & 0 & (A_1+D_1k_z)k_-+F_1k_zk^2_+ & i(A_1+D_1k_z)k_+ +iF_1k_zk^2_- \\
0 & -B_1k_z & i(A_2-D_2k_z)k_-  +iF_2k_zk^2_+ &  (-A_2-D_1k_z) k_+ +F_2k_zk^2_- \\
(A_1+D_1k_z)k_+ +F_1k_zk^2_-  &-i(A_2-D_2k_z)k_+  -iF_2k_zk^2_-  & 0 & iA_3 k_-\\
-i(A_1+D_1k_z)k_- -iF_1k_zk^2_+ &  (-A_2-D_1k_z) k_- +F_2k_zk^2_+ &  -iA_3 k_+ &  0
\end{array}
\right) .$
\end{adjustbox}
\end{align}
\end{widetext}
The double degenerate $E_{3/2}$ band splits into two non-degenerate bands while the double degenerate $E_{5/2}$ band is not affected. Eventually, the Dirac points transform into two pairs of triple nodal points. 

%D2h
{$\boldsymbol{D_{2h}}$}.  ${D_{2h}}$ is the subgroup of $D_{6h}$ obtained by breaking the $C_3$ rotational symmetry. The vertical mirror symmetries $\sigma_x$ and $\sigma_y$, the horizontal mirror symmetry $\sigma_z$ and inversion symmetry are preserved. The corresponding $k\cdot p$ model is 

\begin{widetext}
\begin{align}
\begin{adjustbox}{max width=\textwidth}
$H'_{D_{2h}}({\bf k})=
\left(
\begin{array}{cccc}
0 & 0 & Bk_z & A_1k_- + A_2 k_+ \\
0 & 0 & A_1k_+ + A_2 k_- &  -Bk_z \\
Bk_z  & A_1k_- + A_2 k_+ & 0 & 0 \\
A_1k_+ + A_2 k_-& -Bk_z & 0 & 0
\end{array}
\right) .$
\end{adjustbox}
\end{align}
\end{widetext}
Since there are many high-order terms of $k$, here we only list the linear terms of $k$. The $PT$ symmetry protects the double degeneracy of each band. However, the $C_3$ symmetry breaking introduces the $Bk_z$ hybridization term between the $E_{3/2}$and $E_{5/2}$ bands. Such hybridization will open a gap at the Dirac point, thus transforming the Dirac semimetal into a topological insulator.

%C2v
{$\boldsymbol{C_{2v}}$}.  ${C_{2v}}$ is the subgroup of $D_{2h}$ obtained by breaking inversion symmetry. Vertical mirror symmetries $\sigma_x$ and $\sigma_y$ are preserved. The corresponding $k\cdot p$ model is 

\begin{widetext}
\begin{align}
\begin{adjustbox}{max width=\textwidth}
$H'_{D_{2v}}({\bf k})=
\left(
\begin{array}{cccc}
0 & iB_1k_++iB_2k_- & iBk_z+\Delta & iA_1k_- + iA_2 k_+ \\
 -iB_1k_--iB_2k_+ & 0 & -iA_1k_+ -i A_2 k_- &  iBk_z+\Delta \\
- iBk_z+\Delta  & iA_1k_- + iA_2 k_+ & 0 &  iB'_1k_++iB'_2k_- \\
-iA_1k_+ -i A_2 k_-& - iBk_z+\Delta &  -iB'_1k_--iB'_2k_+ & 0
\end{array}
\right) .$
\end{adjustbox}
\end{align}
\end{widetext}
There is an additional constant hybridization term $\Delta$ between the $E_{3/2}$ and $E_{5/2}$ bands as in the case of $D_{2h}$ group. The gap of the Dirac point will be opened, leading to the strong topological insulator phase.

%C2 or Cs
{$\boldsymbol{C_{2}}$}.  ${C_{2}}$ is the subgroup of $C_{2v}$ obtained by breaking the mirror symmetry but leaving the two-fold rotational symmetry $C_{2z}$. The corresponding $k\cdot p$ model is 

\begin{widetext}
\begin{align}
\begin{adjustbox}{max width=\textwidth}
$H'_{D_{2v}}({\bf k})=
\left(
\begin{array}{cccc}
B_3k_z & iB_1k_++iB_2k_- & iBk_z+\Delta & iA_1k_- + iA_2 k_+ \\
 -iB_1k_--iB_2k_+ & -B_3k_z & -iA_1k_+ -i A_2 k_- &  iBk_z+\Delta \\
- iBk_z+\Delta  & iA_1k_- + iA_2 k_+ & B'_3k_z &  iB'_1k_++iB'_2k_- \\
-iA_1k_+ -i A_2 k_-& - iBk_z+\Delta &  -iB'_1k_--iB'_2k_+ & -B'_3k_z
\end{array}
\right) .$
\end{adjustbox}
\end{align}
\end{widetext}
The breaking of the mirror symmetries $\sigma_x$ and $\sigma_y$ allows the presence of the $B_3k_z\sigma_z$ term in the diagonal part. The $B_3k_z\sigma_z$ term lifts the double degeneracies of $E_{3/2}$ and $E_{5/2}$ bands. Accidental Weyl points could be realized in this little group. 

%section
\section{\label{appendix:kpsurface} $k\cdot p$ model for the (010) surface at the $\bar{\Gamma}$ point}
As mentioned in the main text, the space group of the (010) surface is Pma2 (No. 28), whose generators are the 2-fold rotational symmetry $C_{2y}$ with axis along the $y$ direction and the $G_x=\{\sigma_x|(0,0,c/2)\}$ glide symmetry. Here, we adopt the same coordinates as in the case of the bulk system. The $x$ axis is parallel to the {\bf a} axis. The $y$ axis is parallel to the [120] direction. The $z$ axis is parallel to the {\bf c} axis. On the (010) surface, $k_y$ is not a good quantum number, while $k_x$ and $k_z$ are good quantum numbers. For the $C_{2v}$ point group, there is only one irreducible representation $E_{1/2}$ in the presence of SOC. Hence, the $k\cdot p$ model at the $\Gamma$ point is
\begin{align}
H_{(010)}({\bf \bar{k}})= C_0+ C_1\bar{k}_x^2+ C_2\bar{k}_z^2+\bar{k}_x\sigma_y+\bar{k}_z\sigma_x
\end{align}
At the $\Gamma$ point, the bands are double degenerate. Along the $\bar{k}_z$ direction, $\bar{k}_z\sigma_x$ lifts the degeneracy. As a results the surface energy bands split along $\bar{\Gamma}-\bar{A}$ as shown in Fig.~\ref{fig3}(f) of the main text.

\bibliography{paper}

\begin{thebibliography}{52}%
\makeatletter
\providecommand \@ifxundefined [1]{%
 \@ifx{#1\undefined}
}%
\providecommand \@ifnum [1]{%
 \ifnum #1\expandafter \@firstoftwo
 \else \expandafter \@secondoftwo
 \fi
}%
\providecommand \@ifx [1]{%
 \ifx #1\expandafter \@firstoftwo
 \else \expandafter \@secondoftwo
 \fi
}%
\providecommand \natexlab [1]{#1}%
\providecommand \enquote  [1]{``#1''}%
\providecommand \bibnamefont  [1]{#1}%
\providecommand \bibfnamefont [1]{#1}%
\providecommand \citenamefont [1]{#1}%
\providecommand \href@noop [0]{\@secondoftwo}%
\providecommand \href [0]{\begingroup \@sanitize@url \@href}%
\providecommand \@href[1]{\@@startlink{#1}\@@href}%
\providecommand \@@href[1]{\endgroup#1\@@endlink}%
\providecommand \@sanitize@url [0]{\catcode `\\12\catcode `\$12\catcode
  `\&12\catcode `\#12\catcode `\^12\catcode `\_12\catcode `\%12\relax}%
\providecommand \@@startlink[1]{}%
\providecommand \@@endlink[0]{}%
\providecommand \url  [0]{\begingroup\@sanitize@url \@url }%
\providecommand \@url [1]{\endgroup\@href {#1}{\urlprefix }}%
\providecommand \urlprefix  [0]{URL }%
\providecommand \Eprint [0]{\href }%
\providecommand \doibase [0]{http://dx.doi.org/}%
\providecommand \selectlanguage [0]{\@gobble}%
\providecommand \bibinfo  [0]{\@secondoftwo}%
\providecommand \bibfield  [0]{\@secondoftwo}%
\providecommand \translation [1]{[#1]}%
\providecommand \BibitemOpen [0]{}%
\providecommand \bibitemStop [0]{}%
\providecommand \bibitemNoStop [0]{.\EOS\space}%
\providecommand \EOS [0]{\spacefactor3000\relax}%
\providecommand \BibitemShut  [1]{\csname bibitem#1\endcsname}%
\let\auto@bib@innerbib\@empty
%</preamble>
\bibitem [{\citenamefont {Young}\ \emph {et~al.}(2012)\citenamefont {Young},
  \citenamefont {Zaheer}, \citenamefont {Teo}, \citenamefont {Kane},
  \citenamefont {Mele},\ and\ \citenamefont {Rappe}}]{PhysRevLett.108.140405}%
  \BibitemOpen
  \bibfield  {author} {\bibinfo {author} {\bibfnamefont {S.~M.}\ \bibnamefont
  {Young}}, \bibinfo {author} {\bibfnamefont {S.}~\bibnamefont {Zaheer}},
  \bibinfo {author} {\bibfnamefont {J.~C.~Y.}\ \bibnamefont {Teo}}, \bibinfo
  {author} {\bibfnamefont {C.~L.}\ \bibnamefont {Kane}}, \bibinfo {author}
  {\bibfnamefont {E.~J.}\ \bibnamefont {Mele}}, \ and\ \bibinfo {author}
  {\bibfnamefont {A.~M.}\ \bibnamefont {Rappe}},\ }\href {\doibase
  10.1103/PhysRevLett.108.140405} {\bibfield  {journal} {\bibinfo  {journal}
  {Phys. Rev. Lett.}\ }\textbf {\bibinfo {volume} {108}},\ \bibinfo {pages}
  {140405} (\bibinfo {year} {2012})}\BibitemShut {NoStop}%
\bibitem [{\citenamefont {Wang}\ \emph {et~al.}(2012)\citenamefont {Wang},
  \citenamefont {Sun}, \citenamefont {Chen}, \citenamefont {Franchini},
  \citenamefont {Xu}, \citenamefont {Weng}, \citenamefont {Dai},\ and\
  \citenamefont {Fang}}]{PhysRevB.85.195320}%
  \BibitemOpen
  \bibfield  {author} {\bibinfo {author} {\bibfnamefont {Z.}~\bibnamefont
  {Wang}}, \bibinfo {author} {\bibfnamefont {Y.}~\bibnamefont {Sun}}, \bibinfo
  {author} {\bibfnamefont {X.-Q.}\ \bibnamefont {Chen}}, \bibinfo {author}
  {\bibfnamefont {C.}~\bibnamefont {Franchini}}, \bibinfo {author}
  {\bibfnamefont {G.}~\bibnamefont {Xu}}, \bibinfo {author} {\bibfnamefont
  {H.}~\bibnamefont {Weng}}, \bibinfo {author} {\bibfnamefont {X.}~\bibnamefont
  {Dai}}, \ and\ \bibinfo {author} {\bibfnamefont {Z.}~\bibnamefont {Fang}},\
  }\href {\doibase 10.1103/PhysRevB.85.195320} {\bibfield  {journal} {\bibinfo
  {journal} {Phys. Rev. B}\ }\textbf {\bibinfo {volume} {85}},\ \bibinfo
  {pages} {195320} (\bibinfo {year} {2012})}\BibitemShut {NoStop}%
\bibitem [{\citenamefont {Wang}\ \emph {et~al.}(2013)\citenamefont {Wang},
  \citenamefont {Weng}, \citenamefont {Wu}, \citenamefont {Dai},\ and\
  \citenamefont {Fang}}]{PhysRevB.88.125427}%
  \BibitemOpen
  \bibfield  {author} {\bibinfo {author} {\bibfnamefont {Z.}~\bibnamefont
  {Wang}}, \bibinfo {author} {\bibfnamefont {H.}~\bibnamefont {Weng}}, \bibinfo
  {author} {\bibfnamefont {Q.}~\bibnamefont {Wu}}, \bibinfo {author}
  {\bibfnamefont {X.}~\bibnamefont {Dai}}, \ and\ \bibinfo {author}
  {\bibfnamefont {Z.}~\bibnamefont {Fang}},\ }\href {\doibase
  10.1103/PhysRevB.88.125427} {\bibfield  {journal} {\bibinfo  {journal} {Phys.
  Rev. B}\ }\textbf {\bibinfo {volume} {88}},\ \bibinfo {pages} {125427}
  (\bibinfo {year} {2013})}\BibitemShut {NoStop}%
\bibitem [{\citenamefont {Xiong}\ \emph {et~al.}(2015)\citenamefont {Xiong},
  \citenamefont {Kushwaha}, \citenamefont {Liang}, \citenamefont {Krizan},
  \citenamefont {Hirschberger}, \citenamefont {Wang}, \citenamefont {Cava},\
  and\ \citenamefont {Ong}}]{Xiong413}%
  \BibitemOpen
  \bibfield  {author} {\bibinfo {author} {\bibfnamefont {J.}~\bibnamefont
  {Xiong}}, \bibinfo {author} {\bibfnamefont {S.~K.}\ \bibnamefont {Kushwaha}},
  \bibinfo {author} {\bibfnamefont {T.}~\bibnamefont {Liang}}, \bibinfo
  {author} {\bibfnamefont {J.~W.}\ \bibnamefont {Krizan}}, \bibinfo {author}
  {\bibfnamefont {M.}~\bibnamefont {Hirschberger}}, \bibinfo {author}
  {\bibfnamefont {W.}~\bibnamefont {Wang}}, \bibinfo {author} {\bibfnamefont
  {R.~J.}\ \bibnamefont {Cava}}, \ and\ \bibinfo {author} {\bibfnamefont
  {N.~P.}\ \bibnamefont {Ong}},\ }\href {\doibase 10.1126/science.aac6089}
  {\bibfield  {journal} {\bibinfo  {journal} {Science}\ }\textbf {\bibinfo
  {volume} {350}},\ \bibinfo {pages} {413} (\bibinfo {year}
  {2015})}\BibitemShut {NoStop}%
\bibitem [{\citenamefont {Burkov}(2018)}]{PhysRevLett.120.016603}%
  \BibitemOpen
  \bibfield  {author} {\bibinfo {author} {\bibfnamefont {A.~A.}\ \bibnamefont
  {Burkov}},\ }\href {\doibase 10.1103/PhysRevLett.120.016603} {\bibfield
  {journal} {\bibinfo  {journal} {Phys. Rev. Lett.}\ }\textbf {\bibinfo
  {volume} {120}},\ \bibinfo {pages} {016603} (\bibinfo {year}
  {2018})}\BibitemShut {NoStop}%
\bibitem [{\citenamefont {Yang}\ \emph {et~al.}(2015)\citenamefont {Yang},
  \citenamefont {Morimoto},\ and\ \citenamefont
  {Furusaki}}]{PhysRevB.92.165120}%
  \BibitemOpen
  \bibfield  {author} {\bibinfo {author} {\bibfnamefont {B.-J.}\ \bibnamefont
  {Yang}}, \bibinfo {author} {\bibfnamefont {T.}~\bibnamefont {Morimoto}}, \
  and\ \bibinfo {author} {\bibfnamefont {A.}~\bibnamefont {Furusaki}},\ }\href
  {\doibase 10.1103/PhysRevB.92.165120} {\bibfield  {journal} {\bibinfo
  {journal} {Phys. Rev. B}\ }\textbf {\bibinfo {volume} {92}},\ \bibinfo
  {pages} {165120} (\bibinfo {year} {2015})}\BibitemShut {NoStop}%
\bibitem [{\citenamefont {Armitage}\ \emph {et~al.}(2018)\citenamefont
  {Armitage}, \citenamefont {Mele},\ and\ \citenamefont
  {Vishwanath}}]{RevModPhys.90.015001}%
  \BibitemOpen
  \bibfield  {author} {\bibinfo {author} {\bibfnamefont {N.~P.}\ \bibnamefont
  {Armitage}}, \bibinfo {author} {\bibfnamefont {E.~J.}\ \bibnamefont {Mele}},
  \ and\ \bibinfo {author} {\bibfnamefont {A.}~\bibnamefont {Vishwanath}},\
  }\href {\doibase 10.1103/RevModPhys.90.015001} {\bibfield  {journal}
  {\bibinfo  {journal} {Rev. Mod. Phys.}\ }\textbf {\bibinfo {volume} {90}},\
  \bibinfo {pages} {015001} (\bibinfo {year} {2018})}\BibitemShut {NoStop}%
\bibitem [{\citenamefont {Du}\ \emph {et~al.}(2015)\citenamefont {Du},
  \citenamefont {Wan}, \citenamefont {Wang}, \citenamefont {Sheng},
  \citenamefont {Duan},\ and\ \citenamefont {Wan}}]{Du2015}%
  \BibitemOpen
  \bibfield  {author} {\bibinfo {author} {\bibfnamefont {Y.}~\bibnamefont
  {Du}}, \bibinfo {author} {\bibfnamefont {B.}~\bibnamefont {Wan}}, \bibinfo
  {author} {\bibfnamefont {D.}~\bibnamefont {Wang}}, \bibinfo {author}
  {\bibfnamefont {L.}~\bibnamefont {Sheng}}, \bibinfo {author} {\bibfnamefont
  {C.-G.}\ \bibnamefont {Duan}}, \ and\ \bibinfo {author} {\bibfnamefont
  {X.}~\bibnamefont {Wan}},\ }\href {http://dx.doi.org/10.1038/srep14423}
  {\bibfield  {journal} {\bibinfo  {journal} {Scientific Reports}\ }\textbf
  {\bibinfo {volume} {5}},\ \bibinfo {pages} {14423} (\bibinfo {year}
  {2015})}\BibitemShut {NoStop}%
\bibitem [{\citenamefont {Chen}\ \emph {et~al.}(2017)\citenamefont {Chen},
  \citenamefont {Wang}, \citenamefont {Liu}, \citenamefont {Yu}, \citenamefont
  {Sheng}, \citenamefont {Chen},\ and\ \citenamefont
  {Yang}}]{PhysRevMaterials.1.044201}%
  \BibitemOpen
  \bibfield  {author} {\bibinfo {author} {\bibfnamefont {C.}~\bibnamefont
  {Chen}}, \bibinfo {author} {\bibfnamefont {S.-S.}\ \bibnamefont {Wang}},
  \bibinfo {author} {\bibfnamefont {L.}~\bibnamefont {Liu}}, \bibinfo {author}
  {\bibfnamefont {Z.-M.}\ \bibnamefont {Yu}}, \bibinfo {author} {\bibfnamefont
  {X.-L.}\ \bibnamefont {Sheng}}, \bibinfo {author} {\bibfnamefont
  {Z.}~\bibnamefont {Chen}}, \ and\ \bibinfo {author} {\bibfnamefont {S.~A.}\
  \bibnamefont {Yang}},\ }\href {\doibase 10.1103/PhysRevMaterials.1.044201}
  {\bibfield  {journal} {\bibinfo  {journal} {Phys. Rev. Materials}\ }\textbf
  {\bibinfo {volume} {1}},\ \bibinfo {pages} {044201} (\bibinfo {year}
  {2017})}\BibitemShut {NoStop}%
\bibitem [{\citenamefont {Le}\ \emph {et~al.}(2018)\citenamefont {Le},
  \citenamefont {Wu}, \citenamefont {Qin}, \citenamefont {Li}, \citenamefont
  {Thomale}, \citenamefont {Zhang},\ and\ \citenamefont {Hu}}]{Le2018}%
  \BibitemOpen
  \bibfield  {author} {\bibinfo {author} {\bibfnamefont {C.}~\bibnamefont
  {Le}}, \bibinfo {author} {\bibfnamefont {X.}~\bibnamefont {Wu}}, \bibinfo
  {author} {\bibfnamefont {S.}~\bibnamefont {Qin}}, \bibinfo {author}
  {\bibfnamefont {Y.}~\bibnamefont {Li}}, \bibinfo {author} {\bibfnamefont
  {R.}~\bibnamefont {Thomale}}, \bibinfo {author} {\bibfnamefont
  {F.}~\bibnamefont {Zhang}}, \ and\ \bibinfo {author} {\bibfnamefont
  {J.}~\bibnamefont {Hu}},\ }\href {http://arxiv.org/abs/1801.05719} {\ \textbf
  {\bibinfo {volume} {2}} (\bibinfo {year} {2018})},\ \Eprint
  {http://arxiv.org/abs/1801.05719} {arXiv:1801.05719} \BibitemShut {NoStop}%
\bibitem [{\citenamefont {Liu}\ \emph {et~al.}(2014{\natexlab{a}})\citenamefont
  {Liu}, \citenamefont {Zhou}, \citenamefont {Zhang}, \citenamefont {Wang},
  \citenamefont {Weng}, \citenamefont {Prabhakaran}, \citenamefont {Mo},
  \citenamefont {Shen}, \citenamefont {Fang}, \citenamefont {Dai},
  \citenamefont {Hussain},\ and\ \citenamefont {Chen}}]{Liu864}%
  \BibitemOpen
  \bibfield  {author} {\bibinfo {author} {\bibfnamefont {Z.~K.}\ \bibnamefont
  {Liu}}, \bibinfo {author} {\bibfnamefont {B.}~\bibnamefont {Zhou}}, \bibinfo
  {author} {\bibfnamefont {Y.}~\bibnamefont {Zhang}}, \bibinfo {author}
  {\bibfnamefont {Z.~J.}\ \bibnamefont {Wang}}, \bibinfo {author}
  {\bibfnamefont {H.~M.}\ \bibnamefont {Weng}}, \bibinfo {author}
  {\bibfnamefont {D.}~\bibnamefont {Prabhakaran}}, \bibinfo {author}
  {\bibfnamefont {S.-K.}\ \bibnamefont {Mo}}, \bibinfo {author} {\bibfnamefont
  {Z.~X.}\ \bibnamefont {Shen}}, \bibinfo {author} {\bibfnamefont
  {Z.}~\bibnamefont {Fang}}, \bibinfo {author} {\bibfnamefont {X.}~\bibnamefont
  {Dai}}, \bibinfo {author} {\bibfnamefont {Z.}~\bibnamefont {Hussain}}, \ and\
  \bibinfo {author} {\bibfnamefont {Y.~L.}\ \bibnamefont {Chen}},\ }\href
  {\doibase 10.1126/science.1245085} {\bibfield  {journal} {\bibinfo  {journal}
  {Science}\ }\textbf {\bibinfo {volume} {343}},\ \bibinfo {pages} {864}
  (\bibinfo {year} {2014}{\natexlab{a}})}\BibitemShut {NoStop}%
\bibitem [{\citenamefont {Neupane}\ \emph {et~al.}(2014)\citenamefont
  {Neupane}, \citenamefont {Xu}, \citenamefont {Sankar}, \citenamefont
  {Alidoust}, \citenamefont {Bian}, \citenamefont {Liu}, \citenamefont
  {Belopolski}, \citenamefont {Chang}, \citenamefont {Jeng}, \citenamefont
  {Lin}, \citenamefont {Bansil}, \citenamefont {Chou},\ and\ \citenamefont
  {Hasan}}]{Neupane2014}%
  \BibitemOpen
  \bibfield  {author} {\bibinfo {author} {\bibfnamefont {M.}~\bibnamefont
  {Neupane}}, \bibinfo {author} {\bibfnamefont {S.-Y.}\ \bibnamefont {Xu}},
  \bibinfo {author} {\bibfnamefont {R.}~\bibnamefont {Sankar}}, \bibinfo
  {author} {\bibfnamefont {N.}~\bibnamefont {Alidoust}}, \bibinfo {author}
  {\bibfnamefont {G.}~\bibnamefont {Bian}}, \bibinfo {author} {\bibfnamefont
  {C.}~\bibnamefont {Liu}}, \bibinfo {author} {\bibfnamefont {I.}~\bibnamefont
  {Belopolski}}, \bibinfo {author} {\bibfnamefont {T.-R.}\ \bibnamefont
  {Chang}}, \bibinfo {author} {\bibfnamefont {H.-T.}\ \bibnamefont {Jeng}},
  \bibinfo {author} {\bibfnamefont {H.}~\bibnamefont {Lin}}, \bibinfo {author}
  {\bibfnamefont {A.}~\bibnamefont {Bansil}}, \bibinfo {author} {\bibfnamefont
  {F.}~\bibnamefont {Chou}}, \ and\ \bibinfo {author} {\bibfnamefont {M.~Z.}\
  \bibnamefont {Hasan}},\ }\href {http://dx.doi.org/10.1038/ncomms4786
  http://10.0.4.14/ncomms4786} {\bibfield  {journal} {\bibinfo  {journal}
  {Nature Communications}\ }\textbf {\bibinfo {volume} {5}},\ \bibinfo {pages}
  {3786} (\bibinfo {year} {2014})}\BibitemShut {NoStop}%
\bibitem [{\citenamefont {Brokamp}\ and\ \citenamefont
  {Jacobs}(1992)}]{BROKAMP1992325}%
  \BibitemOpen
  \bibfield  {author} {\bibinfo {author} {\bibfnamefont {T.}~\bibnamefont
  {Brokamp}}\ and\ \bibinfo {author} {\bibfnamefont {H.}~\bibnamefont
  {Jacobs}},\ }\href {\doibase https://doi.org/10.1016/0925-8388(92)90756-Y}
  {\bibfield  {journal} {\bibinfo  {journal} {Journal of Alloys and Compounds}\
  }\textbf {\bibinfo {volume} {183}},\ \bibinfo {pages} {325 } (\bibinfo {year}
  {1992})}\BibitemShut {NoStop}%
\bibitem [{\citenamefont {Miura}\ \emph {et~al.}(2015)\citenamefont {Miura},
  \citenamefont {Tadanaga}, \citenamefont {Magome}, \citenamefont {Moriyoshi},
  \citenamefont {Kuroiwa}, \citenamefont {Takahiro},\ and\ \citenamefont
  {Kumada}}]{ABX2}%
  \BibitemOpen
  \bibfield  {author} {\bibinfo {author} {\bibfnamefont {A.}~\bibnamefont
  {Miura}}, \bibinfo {author} {\bibfnamefont {K.}~\bibnamefont {Tadanaga}},
  \bibinfo {author} {\bibfnamefont {E.}~\bibnamefont {Magome}}, \bibinfo
  {author} {\bibfnamefont {C.}~\bibnamefont {Moriyoshi}}, \bibinfo {author}
  {\bibfnamefont {Y.}~\bibnamefont {Kuroiwa}}, \bibinfo {author} {\bibfnamefont
  {T.}~\bibnamefont {Takahiro}}, \ and\ \bibinfo {author} {\bibfnamefont
  {N.}~\bibnamefont {Kumada}},\ }\href {\doibase
  https://doi.org/10.1016/j.jssc.2015.06.028} {\bibfield  {journal} {\bibinfo
  {journal} {Journal of Solid State Chemistry}\ }\textbf {\bibinfo {volume}
  {229}},\ \bibinfo {pages} {272 } (\bibinfo {year} {2015})}\BibitemShut
  {NoStop}%
\bibitem [{\citenamefont {Niewa}\ \emph {et~al.}(2004)\citenamefont {Niewa},
  \citenamefont {Zherebtsov}, \citenamefont {Schnelle},\ and\ \citenamefont
  {Wagner}}]{ScTaN2}%
  \BibitemOpen
  \bibfield  {author} {\bibinfo {author} {\bibfnamefont {R.}~\bibnamefont
  {Niewa}}, \bibinfo {author} {\bibfnamefont {D.~A.}\ \bibnamefont
  {Zherebtsov}}, \bibinfo {author} {\bibfnamefont {W.}~\bibnamefont
  {Schnelle}}, \ and\ \bibinfo {author} {\bibfnamefont {F.~R.}\ \bibnamefont
  {Wagner}},\ }\href {\doibase 10.1021/ic040027+} {\bibfield  {journal}
  {\bibinfo  {journal} {Inorganic Chemistry}\ }\textbf {\bibinfo {volume}
  {43}},\ \bibinfo {pages} {6188} (\bibinfo {year} {2004})}\BibitemShut
  {NoStop}%
\bibitem [{\citenamefont {Kresse}\ and\ \citenamefont
  {Furthm\"uller}(1996)}]{PhysRevB.54.11169}%
  \BibitemOpen
  \bibfield  {author} {\bibinfo {author} {\bibfnamefont {G.}~\bibnamefont
  {Kresse}}\ and\ \bibinfo {author} {\bibfnamefont {J.}~\bibnamefont
  {Furthm\"uller}},\ }\href {\doibase 10.1103/PhysRevB.54.11169} {\bibfield
  {journal} {\bibinfo  {journal} {Phys. Rev. B}\ }\textbf {\bibinfo {volume}
  {54}},\ \bibinfo {pages} {11169} (\bibinfo {year} {1996})}\BibitemShut
  {NoStop}%
\bibitem [{\citenamefont {Kresse}\ and\ \citenamefont
  {Joubert}(1999)}]{PhysRevB.59.1758}%
  \BibitemOpen
  \bibfield  {author} {\bibinfo {author} {\bibfnamefont {G.}~\bibnamefont
  {Kresse}}\ and\ \bibinfo {author} {\bibfnamefont {D.}~\bibnamefont
  {Joubert}},\ }\href {\doibase 10.1103/PhysRevB.59.1758} {\bibfield  {journal}
  {\bibinfo  {journal} {Phys. Rev. B}\ }\textbf {\bibinfo {volume} {59}},\
  \bibinfo {pages} {1758} (\bibinfo {year} {1999})}\BibitemShut {NoStop}%
\bibitem [{\citenamefont {Bl\"ochl}(1994)}]{PhysRevB.50.17953}%
  \BibitemOpen
  \bibfield  {author} {\bibinfo {author} {\bibfnamefont {P.~E.}\ \bibnamefont
  {Bl\"ochl}},\ }\href {\doibase 10.1103/PhysRevB.50.17953} {\bibfield
  {journal} {\bibinfo  {journal} {Phys. Rev. B}\ }\textbf {\bibinfo {volume}
  {50}},\ \bibinfo {pages} {17953} (\bibinfo {year} {1994})}\BibitemShut
  {NoStop}%
\bibitem [{\citenamefont {Perdew}\ \emph {et~al.}(1996)\citenamefont {Perdew},
  \citenamefont {Burke},\ and\ \citenamefont
  {Ernzerhof}}]{PhysRevLett.77.3865}%
  \BibitemOpen
  \bibfield  {author} {\bibinfo {author} {\bibfnamefont {J.~P.}\ \bibnamefont
  {Perdew}}, \bibinfo {author} {\bibfnamefont {K.}~\bibnamefont {Burke}}, \
  and\ \bibinfo {author} {\bibfnamefont {M.}~\bibnamefont {Ernzerhof}},\ }\href
  {\doibase 10.1103/PhysRevLett.77.3865} {\bibfield  {journal} {\bibinfo
  {journal} {Phys. Rev. Lett.}\ }\textbf {\bibinfo {volume} {77}},\ \bibinfo
  {pages} {3865} (\bibinfo {year} {1996})}\BibitemShut {NoStop}%
\bibitem [{\citenamefont {Heyd}\ \emph {et~al.}(2003)\citenamefont {Heyd},
  \citenamefont {Scuseria},\ and\ \citenamefont {Ernzerhof}}]{HSE06}%
  \BibitemOpen
  \bibfield  {author} {\bibinfo {author} {\bibfnamefont {J.}~\bibnamefont
  {Heyd}}, \bibinfo {author} {\bibfnamefont {G.~E.}\ \bibnamefont {Scuseria}},
  \ and\ \bibinfo {author} {\bibfnamefont {M.}~\bibnamefont {Ernzerhof}},\
  }\href {\doibase 10.1063/1.1564060} {\bibfield  {journal} {\bibinfo
  {journal} {The Journal of Chemical Physics}\ }\textbf {\bibinfo {volume}
  {118}},\ \bibinfo {pages} {8207} (\bibinfo {year} {2003})}\BibitemShut
  {NoStop}%
\bibitem [{SM()}]{SM}%
  \BibitemOpen
  \href@noop {} {}\bibinfo {note} {The supplementary material document contains
  the comparison of band structures obtained using PBE and HSE06 functionals
  and describes effective $k\cdot p$ models at the $\Gamma$ point in presence
  of various symmetry-breaking distortions.}\BibitemShut {Stop}%
\bibitem [{\citenamefont {Wu}\ \emph {et~al.}(2018)\citenamefont {Wu},
  \citenamefont {Zhang}, \citenamefont {Song}, \citenamefont {Troyer},\ and\
  \citenamefont {Soluyanov}}]{wanniertools}%
  \BibitemOpen
  \bibfield  {author} {\bibinfo {author} {\bibfnamefont {Q.}~\bibnamefont
  {Wu}}, \bibinfo {author} {\bibfnamefont {S.}~\bibnamefont {Zhang}}, \bibinfo
  {author} {\bibfnamefont {H.-F.}\ \bibnamefont {Song}}, \bibinfo {author}
  {\bibfnamefont {M.}~\bibnamefont {Troyer}}, \ and\ \bibinfo {author}
  {\bibfnamefont {A.~A.}\ \bibnamefont {Soluyanov}},\ }\href {\doibase
  https://doi.org/10.1016/j.cpc.2017.09.033} {\bibfield  {journal} {\bibinfo
  {journal} {Computer Physics Communications}\ }\textbf {\bibinfo {volume}
  {224}},\ \bibinfo {pages} {405 } (\bibinfo {year} {2018})}\BibitemShut
  {NoStop}%
\bibitem [{\citenamefont {Marzari}\ \emph {et~al.}(2012)\citenamefont
  {Marzari}, \citenamefont {Mostofi}, \citenamefont {Yates}, \citenamefont
  {Souza},\ and\ \citenamefont {Vanderbilt}}]{RevModPhys.84.1419}%
  \BibitemOpen
  \bibfield  {author} {\bibinfo {author} {\bibfnamefont {N.}~\bibnamefont
  {Marzari}}, \bibinfo {author} {\bibfnamefont {A.~A.}\ \bibnamefont
  {Mostofi}}, \bibinfo {author} {\bibfnamefont {J.~R.}\ \bibnamefont {Yates}},
  \bibinfo {author} {\bibfnamefont {I.}~\bibnamefont {Souza}}, \ and\ \bibinfo
  {author} {\bibfnamefont {D.}~\bibnamefont {Vanderbilt}},\ }\href {\doibase
  10.1103/RevModPhys.84.1419} {\bibfield  {journal} {\bibinfo  {journal} {Rev.
  Mod. Phys.}\ }\textbf {\bibinfo {volume} {84}},\ \bibinfo {pages} {1419}
  (\bibinfo {year} {2012})}\BibitemShut {NoStop}%
\bibitem [{\citenamefont {Mostofi}\ \emph {et~al.}(2014)\citenamefont
  {Mostofi}, \citenamefont {Yates}, \citenamefont {Pizzi}, \citenamefont {Lee},
  \citenamefont {Souza}, \citenamefont {Vanderbilt},\ and\ \citenamefont
  {Marzari}}]{wannier90}%
  \BibitemOpen
  \bibfield  {author} {\bibinfo {author} {\bibfnamefont {A.~A.}\ \bibnamefont
  {Mostofi}}, \bibinfo {author} {\bibfnamefont {J.~R.}\ \bibnamefont {Yates}},
  \bibinfo {author} {\bibfnamefont {G.}~\bibnamefont {Pizzi}}, \bibinfo
  {author} {\bibfnamefont {Y.-S.}\ \bibnamefont {Lee}}, \bibinfo {author}
  {\bibfnamefont {I.}~\bibnamefont {Souza}}, \bibinfo {author} {\bibfnamefont
  {D.}~\bibnamefont {Vanderbilt}}, \ and\ \bibinfo {author} {\bibfnamefont
  {N.}~\bibnamefont {Marzari}},\ }\href {\doibase
  https://doi.org/10.1016/j.cpc.2014.05.003} {\bibfield  {journal} {\bibinfo
  {journal} {Computer Physics Communications}\ }\textbf {\bibinfo {volume}
  {185}},\ \bibinfo {pages} {2309 } (\bibinfo {year} {2014})}\BibitemShut
  {NoStop}%
\bibitem [{\citenamefont {Sancho}\ \emph {et~al.}(1985)\citenamefont {Sancho},
  \citenamefont {Sancho}, \citenamefont {Sancho},\ and\ \citenamefont
  {Rubio}}]{0305-4608-15-4-009}%
  \BibitemOpen
  \bibfield  {author} {\bibinfo {author} {\bibfnamefont {M.~P.~L.}\
  \bibnamefont {Sancho}}, \bibinfo {author} {\bibfnamefont {J.~M.~L.}\
  \bibnamefont {Sancho}}, \bibinfo {author} {\bibfnamefont {J.~M.~L.}\
  \bibnamefont {Sancho}}, \ and\ \bibinfo {author} {\bibfnamefont
  {J.}~\bibnamefont {Rubio}},\ }\href
  {http://stacks.iop.org/0305-4608/15/i=4/a=009} {\bibfield  {journal}
  {\bibinfo  {journal} {Journal of Physics F: Metal Physics}\ }\textbf
  {\bibinfo {volume} {15}},\ \bibinfo {pages} {851} (\bibinfo {year}
  {1985})}\BibitemShut {NoStop}%
\bibitem [{\citenamefont {Altmann}\ and\ \citenamefont
  {Herzig}(1994{\natexlab{a}})}]{altmann1994}%
  \BibitemOpen
  \bibfield  {author} {\bibinfo {author} {\bibfnamefont {S.}~\bibnamefont
  {Altmann}}\ and\ \bibinfo {author} {\bibfnamefont {P.}~\bibnamefont
  {Herzig}},\ }\href@noop {} {\enquote {\bibinfo {title} {Point-group theory
  tables},}\ }\bibinfo {howpublished} {Clarendon Press} (\bibinfo {year}
  {1994}{\natexlab{a}}),\ \bibinfo {note} {the point group citations: $D_{6h}$
  in page 273, $C_{6v}$ in page 497, $D_{3d}$ in page 370, $D_{3h}$ in page
  250, $C_{3v}$ in page 484, $D_6$ in page 207, $C_6$ in page 119, $S_6$ in
  page 146, $D_3$ in page 196, $C_3$ in page 112, $D_{2h}$ in page 246,
  $C_{2v}$ in page 482, $C_s$ in page 140, $C_2$ in page 110. The subduction
  table for $D_{6h}$ is T 35.9 in page 282.}\BibitemShut {Stop}%
\bibitem [{\citenamefont {Jenkins}\ \emph {et~al.}(2016)\citenamefont
  {Jenkins}, \citenamefont {Lane}, \citenamefont {Barbiellini}, \citenamefont
  {Sushkov}, \citenamefont {Carey}, \citenamefont {Liu}, \citenamefont
  {Krizan}, \citenamefont {Kushwaha}, \citenamefont {Gibson}, \citenamefont
  {Chang}, \citenamefont {Jeng}, \citenamefont {Lin}, \citenamefont {Cava},
  \citenamefont {Bansil},\ and\ \citenamefont {Drew}}]{PhysRevB.94.085121}%
  \BibitemOpen
  \bibfield  {author} {\bibinfo {author} {\bibfnamefont {G.~S.}\ \bibnamefont
  {Jenkins}}, \bibinfo {author} {\bibfnamefont {C.}~\bibnamefont {Lane}},
  \bibinfo {author} {\bibfnamefont {B.}~\bibnamefont {Barbiellini}}, \bibinfo
  {author} {\bibfnamefont {A.~B.}\ \bibnamefont {Sushkov}}, \bibinfo {author}
  {\bibfnamefont {R.~L.}\ \bibnamefont {Carey}}, \bibinfo {author}
  {\bibfnamefont {F.}~\bibnamefont {Liu}}, \bibinfo {author} {\bibfnamefont
  {J.~W.}\ \bibnamefont {Krizan}}, \bibinfo {author} {\bibfnamefont {S.~K.}\
  \bibnamefont {Kushwaha}}, \bibinfo {author} {\bibfnamefont {Q.}~\bibnamefont
  {Gibson}}, \bibinfo {author} {\bibfnamefont {T.-R.}\ \bibnamefont {Chang}},
  \bibinfo {author} {\bibfnamefont {H.-T.}\ \bibnamefont {Jeng}}, \bibinfo
  {author} {\bibfnamefont {H.}~\bibnamefont {Lin}}, \bibinfo {author}
  {\bibfnamefont {R.~J.}\ \bibnamefont {Cava}}, \bibinfo {author}
  {\bibfnamefont {A.}~\bibnamefont {Bansil}}, \ and\ \bibinfo {author}
  {\bibfnamefont {H.~D.}\ \bibnamefont {Drew}},\ }\href {\doibase
  10.1103/PhysRevB.94.085121} {\bibfield  {journal} {\bibinfo  {journal} {Phys.
  Rev. B}\ }\textbf {\bibinfo {volume} {94}},\ \bibinfo {pages} {085121}
  (\bibinfo {year} {2016})}\BibitemShut {NoStop}%
\bibitem [{\citenamefont {Kushwaha}\ \emph {et~al.}(2015)\citenamefont
  {Kushwaha}, \citenamefont {Krizan}, \citenamefont {Feldman}, \citenamefont
  {Gyenis}, \citenamefont {Randeria}, \citenamefont {Xiong}, \citenamefont
  {Xu}, \citenamefont {Alidoust}, \citenamefont {Belopolski}, \citenamefont
  {Liang}, \citenamefont {Zahid~Hasan}, \citenamefont {Ong}, \citenamefont
  {Yazdani},\ and\ \citenamefont {Cava}}]{Kushwaha2015}%
  \BibitemOpen
  \bibfield  {author} {\bibinfo {author} {\bibfnamefont {S.~K.}\ \bibnamefont
  {Kushwaha}}, \bibinfo {author} {\bibfnamefont {J.~W.}\ \bibnamefont
  {Krizan}}, \bibinfo {author} {\bibfnamefont {B.~E.}\ \bibnamefont {Feldman}},
  \bibinfo {author} {\bibfnamefont {A.}~\bibnamefont {Gyenis}}, \bibinfo
  {author} {\bibfnamefont {M.~T.}\ \bibnamefont {Randeria}}, \bibinfo {author}
  {\bibfnamefont {J.}~\bibnamefont {Xiong}}, \bibinfo {author} {\bibfnamefont
  {S.-Y.}\ \bibnamefont {Xu}}, \bibinfo {author} {\bibfnamefont
  {N.}~\bibnamefont {Alidoust}}, \bibinfo {author} {\bibfnamefont
  {I.}~\bibnamefont {Belopolski}}, \bibinfo {author} {\bibfnamefont
  {T.}~\bibnamefont {Liang}}, \bibinfo {author} {\bibfnamefont
  {M.}~\bibnamefont {Zahid~Hasan}}, \bibinfo {author} {\bibfnamefont {N.~P.}\
  \bibnamefont {Ong}}, \bibinfo {author} {\bibfnamefont {A.}~\bibnamefont
  {Yazdani}}, \ and\ \bibinfo {author} {\bibfnamefont {R.~J.}\ \bibnamefont
  {Cava}},\ }\href {\doibase 10.1063/1.4908158} {\bibfield  {journal} {\bibinfo
   {journal} {APL Materials}\ }\textbf {\bibinfo {volume} {3}},\ \bibinfo
  {pages} {041504} (\bibinfo {year} {2015})}\BibitemShut {NoStop}%
\bibitem [{\citenamefont {Liu}\ \emph {et~al.}(2014{\natexlab{b}})\citenamefont
  {Liu}, \citenamefont {Jiang}, \citenamefont {Zhou}, \citenamefont {Wang},
  \citenamefont {Zhang}, \citenamefont {Weng}, \citenamefont {Prabhakaran},
  \citenamefont {Mo}, \citenamefont {Peng}, \citenamefont {Dudin},
  \citenamefont {Kim}, \citenamefont {Hoesch}, \citenamefont {Fang},
  \citenamefont {Dai}, \citenamefont {Shen}, \citenamefont {Feng},
  \citenamefont {Hussain},\ and\ \citenamefont {Chen}}]{Liu2014}%
  \BibitemOpen
  \bibfield  {author} {\bibinfo {author} {\bibfnamefont {Z.~K.}\ \bibnamefont
  {Liu}}, \bibinfo {author} {\bibfnamefont {J.}~\bibnamefont {Jiang}}, \bibinfo
  {author} {\bibfnamefont {B.}~\bibnamefont {Zhou}}, \bibinfo {author}
  {\bibfnamefont {Z.~J.}\ \bibnamefont {Wang}}, \bibinfo {author}
  {\bibfnamefont {Y.}~\bibnamefont {Zhang}}, \bibinfo {author} {\bibfnamefont
  {H.~M.}\ \bibnamefont {Weng}}, \bibinfo {author} {\bibfnamefont
  {D.}~\bibnamefont {Prabhakaran}}, \bibinfo {author} {\bibfnamefont {S.-K.}\
  \bibnamefont {Mo}}, \bibinfo {author} {\bibfnamefont {H.}~\bibnamefont
  {Peng}}, \bibinfo {author} {\bibfnamefont {P.}~\bibnamefont {Dudin}},
  \bibinfo {author} {\bibfnamefont {T.}~\bibnamefont {Kim}}, \bibinfo {author}
  {\bibfnamefont {M.}~\bibnamefont {Hoesch}}, \bibinfo {author} {\bibfnamefont
  {Z.}~\bibnamefont {Fang}}, \bibinfo {author} {\bibfnamefont {X.}~\bibnamefont
  {Dai}}, \bibinfo {author} {\bibfnamefont {Z.~X.}\ \bibnamefont {Shen}},
  \bibinfo {author} {\bibfnamefont {D.~L.}\ \bibnamefont {Feng}}, \bibinfo
  {author} {\bibfnamefont {Z.}~\bibnamefont {Hussain}}, \ and\ \bibinfo
  {author} {\bibfnamefont {Y.~L.}\ \bibnamefont {Chen}},\ }\href
  {http://dx.doi.org/10.1038/nmat3990 http://10.0.4.14/nmat3990
  https://www.nature.com/articles/nmat3990{\#}supplementary-information}
  {\bibfield  {journal} {\bibinfo  {journal} {Nature Materials}\ }\textbf
  {\bibinfo {volume} {13}},\ \bibinfo {pages} {677} (\bibinfo {year}
  {2014}{\natexlab{b}})}\BibitemShut {NoStop}%
\bibitem [{\citenamefont {Liang}\ \emph {et~al.}(2014)\citenamefont {Liang},
  \citenamefont {Gibson}, \citenamefont {Ali}, \citenamefont {Liu},
  \citenamefont {Cava},\ and\ \citenamefont {Ong}}]{Liang2014}%
  \BibitemOpen
  \bibfield  {author} {\bibinfo {author} {\bibfnamefont {T.}~\bibnamefont
  {Liang}}, \bibinfo {author} {\bibfnamefont {Q.}~\bibnamefont {Gibson}},
  \bibinfo {author} {\bibfnamefont {M.~N.}\ \bibnamefont {Ali}}, \bibinfo
  {author} {\bibfnamefont {M.}~\bibnamefont {Liu}}, \bibinfo {author}
  {\bibfnamefont {R.~J.}\ \bibnamefont {Cava}}, \ and\ \bibinfo {author}
  {\bibfnamefont {N.~P.}\ \bibnamefont {Ong}},\ }\href
  {http://dx.doi.org/10.1038/nmat4143 http://10.0.4.14/nmat4143} {\bibfield
  {journal} {\bibinfo  {journal} {Nature Materials}\ }\textbf {\bibinfo
  {volume} {14}},\ \bibinfo {pages} {280} (\bibinfo {year} {2014})}\BibitemShut
  {NoStop}%
\bibitem [{\citenamefont {Winkler}\ \emph {et~al.}(2016)\citenamefont
  {Winkler}, \citenamefont {Wu}, \citenamefont {Troyer}, \citenamefont
  {Krogstrup},\ and\ \citenamefont {Soluyanov}}]{PhysRevLett.117.076403}%
  \BibitemOpen
  \bibfield  {author} {\bibinfo {author} {\bibfnamefont {G.~W.}\ \bibnamefont
  {Winkler}}, \bibinfo {author} {\bibfnamefont {Q.}~\bibnamefont {Wu}},
  \bibinfo {author} {\bibfnamefont {M.}~\bibnamefont {Troyer}}, \bibinfo
  {author} {\bibfnamefont {P.}~\bibnamefont {Krogstrup}}, \ and\ \bibinfo
  {author} {\bibfnamefont {A.~A.}\ \bibnamefont {Soluyanov}},\ }\href {\doibase
  10.1103/PhysRevLett.117.076403} {\bibfield  {journal} {\bibinfo  {journal}
  {Phys. Rev. Lett.}\ }\textbf {\bibinfo {volume} {117}},\ \bibinfo {pages}
  {076403} (\bibinfo {year} {2016})}\BibitemShut {NoStop}%
\bibitem [{\citenamefont {Zhu}\ \emph {et~al.}(2016)\citenamefont {Zhu},
  \citenamefont {Winkler}, \citenamefont {Wu}, \citenamefont {Li},\ and\
  \citenamefont {Soluyanov}}]{PhysRevX.6.031003}%
  \BibitemOpen
  \bibfield  {author} {\bibinfo {author} {\bibfnamefont {Z.}~\bibnamefont
  {Zhu}}, \bibinfo {author} {\bibfnamefont {G.~W.}\ \bibnamefont {Winkler}},
  \bibinfo {author} {\bibfnamefont {Q.}~\bibnamefont {Wu}}, \bibinfo {author}
  {\bibfnamefont {J.}~\bibnamefont {Li}}, \ and\ \bibinfo {author}
  {\bibfnamefont {A.~A.}\ \bibnamefont {Soluyanov}},\ }\href {\doibase
  10.1103/PhysRevX.6.031003} {\bibfield  {journal} {\bibinfo  {journal} {Phys.
  Rev. X}\ }\textbf {\bibinfo {volume} {6}},\ \bibinfo {pages} {031003}
  (\bibinfo {year} {2016})}\BibitemShut {NoStop}%
\bibitem [{\citenamefont {Weng}\ \emph
  {et~al.}(2016{\natexlab{a}})\citenamefont {Weng}, \citenamefont {Fang},
  \citenamefont {Fang},\ and\ \citenamefont {Dai}}]{PhysRevB.93.241202}%
  \BibitemOpen
  \bibfield  {author} {\bibinfo {author} {\bibfnamefont {H.}~\bibnamefont
  {Weng}}, \bibinfo {author} {\bibfnamefont {C.}~\bibnamefont {Fang}}, \bibinfo
  {author} {\bibfnamefont {Z.}~\bibnamefont {Fang}}, \ and\ \bibinfo {author}
  {\bibfnamefont {X.}~\bibnamefont {Dai}},\ }\href {\doibase
  10.1103/PhysRevB.93.241202} {\bibfield  {journal} {\bibinfo  {journal} {Phys.
  Rev. B}\ }\textbf {\bibinfo {volume} {93}},\ \bibinfo {pages} {241202}
  (\bibinfo {year} {2016}{\natexlab{a}})}\BibitemShut {NoStop}%
\bibitem [{\citenamefont {Weng}\ \emph
  {et~al.}(2016{\natexlab{b}})\citenamefont {Weng}, \citenamefont {Fang},
  \citenamefont {Fang},\ and\ \citenamefont {Dai}}]{PhysRevB.94.165201}%
  \BibitemOpen
  \bibfield  {author} {\bibinfo {author} {\bibfnamefont {H.}~\bibnamefont
  {Weng}}, \bibinfo {author} {\bibfnamefont {C.}~\bibnamefont {Fang}}, \bibinfo
  {author} {\bibfnamefont {Z.}~\bibnamefont {Fang}}, \ and\ \bibinfo {author}
  {\bibfnamefont {X.}~\bibnamefont {Dai}},\ }\href {\doibase
  10.1103/PhysRevB.94.165201} {\bibfield  {journal} {\bibinfo  {journal} {Phys.
  Rev. B}\ }\textbf {\bibinfo {volume} {94}},\ \bibinfo {pages} {165201}
  (\bibinfo {year} {2016}{\natexlab{b}})}\BibitemShut {NoStop}%
\bibitem [{\citenamefont {Wang}\ \emph
  {et~al.}(2017{\natexlab{a}})\citenamefont {Wang}, \citenamefont {Sui},
  \citenamefont {Shi}, \citenamefont {Pan}, \citenamefont {Zhang},
  \citenamefont {Liu}, \citenamefont {Wei}, \citenamefont {Yan},\ and\
  \citenamefont {Huang}}]{PhysRevLett.119.256402}%
  \BibitemOpen
  \bibfield  {author} {\bibinfo {author} {\bibfnamefont {J.}~\bibnamefont
  {Wang}}, \bibinfo {author} {\bibfnamefont {X.}~\bibnamefont {Sui}}, \bibinfo
  {author} {\bibfnamefont {W.}~\bibnamefont {Shi}}, \bibinfo {author}
  {\bibfnamefont {J.}~\bibnamefont {Pan}}, \bibinfo {author} {\bibfnamefont
  {S.}~\bibnamefont {Zhang}}, \bibinfo {author} {\bibfnamefont
  {F.}~\bibnamefont {Liu}}, \bibinfo {author} {\bibfnamefont {S.-H.}\
  \bibnamefont {Wei}}, \bibinfo {author} {\bibfnamefont {Q.}~\bibnamefont
  {Yan}}, \ and\ \bibinfo {author} {\bibfnamefont {B.}~\bibnamefont {Huang}},\
  }\href {\doibase 10.1103/PhysRevLett.119.256402} {\bibfield  {journal}
  {\bibinfo  {journal} {Phys. Rev. Lett.}\ }\textbf {\bibinfo {volume} {119}},\
  \bibinfo {pages} {256402} (\bibinfo {year} {2017}{\natexlab{a}})}\BibitemShut
  {NoStop}%
\bibitem [{\citenamefont {Chang}\ \emph {et~al.}(2017)\citenamefont {Chang},
  \citenamefont {Xu}, \citenamefont {Huang}, \citenamefont {Sanchez},
  \citenamefont {Hsu}, \citenamefont {Bian}, \citenamefont {Yu}, \citenamefont
  {Belopolski}, \citenamefont {Alidoust}, \citenamefont {Zheng}, \citenamefont
  {Chang}, \citenamefont {Jeng}, \citenamefont {Yang}, \citenamefont {Neupert},
  \citenamefont {Lin},\ and\ \citenamefont {Hasan}}]{Chang2017}%
  \BibitemOpen
  \bibfield  {author} {\bibinfo {author} {\bibfnamefont {G.}~\bibnamefont
  {Chang}}, \bibinfo {author} {\bibfnamefont {S.-Y.}\ \bibnamefont {Xu}},
  \bibinfo {author} {\bibfnamefont {S.-M.}\ \bibnamefont {Huang}}, \bibinfo
  {author} {\bibfnamefont {D.~S.}\ \bibnamefont {Sanchez}}, \bibinfo {author}
  {\bibfnamefont {C.-H.}\ \bibnamefont {Hsu}}, \bibinfo {author} {\bibfnamefont
  {G.}~\bibnamefont {Bian}}, \bibinfo {author} {\bibfnamefont {Z.-M.}\
  \bibnamefont {Yu}}, \bibinfo {author} {\bibfnamefont {I.}~\bibnamefont
  {Belopolski}}, \bibinfo {author} {\bibfnamefont {N.}~\bibnamefont
  {Alidoust}}, \bibinfo {author} {\bibfnamefont {H.}~\bibnamefont {Zheng}},
  \bibinfo {author} {\bibfnamefont {T.-R.}\ \bibnamefont {Chang}}, \bibinfo
  {author} {\bibfnamefont {H.-T.}\ \bibnamefont {Jeng}}, \bibinfo {author}
  {\bibfnamefont {S.~A.}\ \bibnamefont {Yang}}, \bibinfo {author}
  {\bibfnamefont {T.}~\bibnamefont {Neupert}}, \bibinfo {author} {\bibfnamefont
  {H.}~\bibnamefont {Lin}}, \ and\ \bibinfo {author} {\bibfnamefont {M.~Z.}\
  \bibnamefont {Hasan}},\ }\href {\doibase 10.1038/s41598-017-01523-8}
  {\bibfield  {journal} {\bibinfo  {journal} {Scientific Reports}\ }\textbf
  {\bibinfo {volume} {7}},\ \bibinfo {pages} {1688} (\bibinfo {year}
  {2017})}\BibitemShut {NoStop}%
\bibitem [{\citenamefont {Lv}\ \emph {et~al.}(2017)\citenamefont {Lv},
  \citenamefont {Feng}, \citenamefont {Xu}, \citenamefont {Gao}, \citenamefont
  {Ma}, \citenamefont {Kong}, \citenamefont {Richard}, \citenamefont {Huang},
  \citenamefont {Strocov}, \citenamefont {Fang}, \citenamefont {Weng},
  \citenamefont {Shi}, \citenamefont {Qian},\ and\ \citenamefont
  {Ding}}]{Lv2017}%
  \BibitemOpen
  \bibfield  {author} {\bibinfo {author} {\bibfnamefont {B.~Q.}\ \bibnamefont
  {Lv}}, \bibinfo {author} {\bibfnamefont {Z.-L.}\ \bibnamefont {Feng}},
  \bibinfo {author} {\bibfnamefont {Q.-N.}\ \bibnamefont {Xu}}, \bibinfo
  {author} {\bibfnamefont {X.}~\bibnamefont {Gao}}, \bibinfo {author}
  {\bibfnamefont {J.-Z.}\ \bibnamefont {Ma}}, \bibinfo {author} {\bibfnamefont
  {L.-Y.}\ \bibnamefont {Kong}}, \bibinfo {author} {\bibfnamefont
  {P.}~\bibnamefont {Richard}}, \bibinfo {author} {\bibfnamefont {Y.-B.}\
  \bibnamefont {Huang}}, \bibinfo {author} {\bibfnamefont {V.~N.}\ \bibnamefont
  {Strocov}}, \bibinfo {author} {\bibfnamefont {C.}~\bibnamefont {Fang}},
  \bibinfo {author} {\bibfnamefont {H.-M.}\ \bibnamefont {Weng}}, \bibinfo
  {author} {\bibfnamefont {Y.-G.}\ \bibnamefont {Shi}}, \bibinfo {author}
  {\bibfnamefont {T.}~\bibnamefont {Qian}}, \ and\ \bibinfo {author}
  {\bibfnamefont {H.}~\bibnamefont {Ding}},\ }\href
  {http://dx.doi.org/10.1038/nature22390 http://10.0.4.14/nature22390}
  {\bibfield  {journal} {\bibinfo  {journal} {Nature}\ }\textbf {\bibinfo
  {volume} {546}},\ \bibinfo {pages} {627} (\bibinfo {year}
  {2017})}\BibitemShut {NoStop}%
\bibitem [{\citenamefont {Yu}\ \emph {et~al.}(2011)\citenamefont {Yu},
  \citenamefont {Qi}, \citenamefont {Bernevig}, \citenamefont {Fang},\ and\
  \citenamefont {Dai}}]{PhysRevB.84.075119}%
  \BibitemOpen
  \bibfield  {author} {\bibinfo {author} {\bibfnamefont {R.}~\bibnamefont
  {Yu}}, \bibinfo {author} {\bibfnamefont {X.~L.}\ \bibnamefont {Qi}}, \bibinfo
  {author} {\bibfnamefont {A.}~\bibnamefont {Bernevig}}, \bibinfo {author}
  {\bibfnamefont {Z.}~\bibnamefont {Fang}}, \ and\ \bibinfo {author}
  {\bibfnamefont {X.}~\bibnamefont {Dai}},\ }\href {\doibase
  10.1103/PhysRevB.84.075119} {\bibfield  {journal} {\bibinfo  {journal} {Phys.
  Rev. B}\ }\textbf {\bibinfo {volume} {84}},\ \bibinfo {pages} {075119}
  (\bibinfo {year} {2011})}\BibitemShut {NoStop}%
\bibitem [{\citenamefont {Soluyanov}\ and\ \citenamefont
  {Vanderbilt}(2011)}]{PhysRevB.83.035108}%
  \BibitemOpen
  \bibfield  {author} {\bibinfo {author} {\bibfnamefont {A.~A.}\ \bibnamefont
  {Soluyanov}}\ and\ \bibinfo {author} {\bibfnamefont {D.}~\bibnamefont
  {Vanderbilt}},\ }\href {\doibase 10.1103/PhysRevB.83.035108} {\bibfield
  {journal} {\bibinfo  {journal} {Phys. Rev. B}\ }\textbf {\bibinfo {volume}
  {83}},\ \bibinfo {pages} {035108} (\bibinfo {year} {2011})}\BibitemShut
  {NoStop}%
\bibitem [{\citenamefont {Young}\ and\ \citenamefont
  {Kane}(2015)}]{PhysRevLett.115.126803}%
  \BibitemOpen
  \bibfield  {author} {\bibinfo {author} {\bibfnamefont {S.~M.}\ \bibnamefont
  {Young}}\ and\ \bibinfo {author} {\bibfnamefont {C.~L.}\ \bibnamefont
  {Kane}},\ }\href {\doibase 10.1103/PhysRevLett.115.126803} {\bibfield
  {journal} {\bibinfo  {journal} {Phys. Rev. Lett.}\ }\textbf {\bibinfo
  {volume} {115}},\ \bibinfo {pages} {126803} (\bibinfo {year}
  {2015})}\BibitemShut {NoStop}%
\bibitem [{\citenamefont {Bzdu{\v{s}}ek}\ \emph {et~al.}(2016)\citenamefont
  {Bzdu{\v{s}}ek}, \citenamefont {Wu}, \citenamefont {R{\"{u}}egg},
  \citenamefont {Sigrist},\ and\ \citenamefont {Soluyanov}}]{Bzdusek2016}%
  \BibitemOpen
  \bibfield  {author} {\bibinfo {author} {\bibfnamefont {T.}~\bibnamefont
  {Bzdu{\v{s}}ek}}, \bibinfo {author} {\bibfnamefont {Q.}~\bibnamefont {Wu}},
  \bibinfo {author} {\bibfnamefont {A.}~\bibnamefont {R{\"{u}}egg}}, \bibinfo
  {author} {\bibfnamefont {M.}~\bibnamefont {Sigrist}}, \ and\ \bibinfo
  {author} {\bibfnamefont {A.~A.}\ \bibnamefont {Soluyanov}},\ }\href {\doibase
  10.1038/nature19099} {\bibfield  {journal} {\bibinfo  {journal} {Nature}\
  }\textbf {\bibinfo {volume} {538}},\ \bibinfo {pages} {75} (\bibinfo {year}
  {2016})}\BibitemShut {NoStop}%
\bibitem [{\citenamefont {Wang}\ \emph {et~al.}(2016)\citenamefont {Wang},
  \citenamefont {Alexandradinata}, \citenamefont {Cava},\ and\ \citenamefont
  {Bernevig}}]{Wang2016}%
  \BibitemOpen
  \bibfield  {author} {\bibinfo {author} {\bibfnamefont {Z.}~\bibnamefont
  {Wang}}, \bibinfo {author} {\bibfnamefont {A.}~\bibnamefont
  {Alexandradinata}}, \bibinfo {author} {\bibfnamefont {R.~J.}\ \bibnamefont
  {Cava}}, \ and\ \bibinfo {author} {\bibfnamefont {B.~A.}\ \bibnamefont
  {Bernevig}},\ }\href {\doibase 10.1038/nature17410} {\bibfield  {journal}
  {\bibinfo  {journal} {Nature}\ }\textbf {\bibinfo {volume} {532}},\ \bibinfo
  {pages} {189} (\bibinfo {year} {2016})}\BibitemShut {NoStop}%
\bibitem [{\citenamefont {Wang}\ \emph
  {et~al.}(2017{\natexlab{b}})\citenamefont {Wang}, \citenamefont {Liu},
  \citenamefont {Yu}, \citenamefont {Sheng},\ and\ \citenamefont
  {Yang}}]{Wang2017}%
  \BibitemOpen
  \bibfield  {author} {\bibinfo {author} {\bibfnamefont {S.-S.}\ \bibnamefont
  {Wang}}, \bibinfo {author} {\bibfnamefont {Y.}~\bibnamefont {Liu}}, \bibinfo
  {author} {\bibfnamefont {Z.-M.}\ \bibnamefont {Yu}}, \bibinfo {author}
  {\bibfnamefont {X.-L.}\ \bibnamefont {Sheng}}, \ and\ \bibinfo {author}
  {\bibfnamefont {S.~A.}\ \bibnamefont {Yang}},\ }\href {\doibase
  10.1038/s41467-017-01986-3} {\bibfield  {journal} {\bibinfo  {journal}
  {Nature Communications}\ }\textbf {\bibinfo {volume} {8}},\ \bibinfo {pages}
  {1844} (\bibinfo {year} {2017}{\natexlab{b}})},\ \Eprint
  {http://arxiv.org/abs/1705.01424} {arXiv:1705.01424} \BibitemShut {NoStop}%
\bibitem [{\citenamefont {Kargarian}\ \emph {et~al.}(2016)\citenamefont
  {Kargarian}, \citenamefont {Randeria},\ and\ \citenamefont
  {Lu}}]{Kargarian8648}%
  \BibitemOpen
  \bibfield  {author} {\bibinfo {author} {\bibfnamefont {M.}~\bibnamefont
  {Kargarian}}, \bibinfo {author} {\bibfnamefont {M.}~\bibnamefont {Randeria}},
  \ and\ \bibinfo {author} {\bibfnamefont {Y.-M.}\ \bibnamefont {Lu}},\ }\href
  {\doibase 10.1073/pnas.1524787113} {\bibfield  {journal} {\bibinfo  {journal}
  {Proceedings of the National Academy of Sciences}\ }\textbf {\bibinfo
  {volume} {113}},\ \bibinfo {pages} {8648} (\bibinfo {year}
  {2016})}\BibitemShut {NoStop}%
\bibitem [{\citenamefont {Kargarian}\ \emph {et~al.}(2018)\citenamefont
  {Kargarian}, \citenamefont {Lu},\ and\ \citenamefont
  {Randeria}}]{PhysRevB.97.165129}%
  \BibitemOpen
  \bibfield  {author} {\bibinfo {author} {\bibfnamefont {M.}~\bibnamefont
  {Kargarian}}, \bibinfo {author} {\bibfnamefont {Y.-M.}\ \bibnamefont {Lu}}, \
  and\ \bibinfo {author} {\bibfnamefont {M.}~\bibnamefont {Randeria}},\ }\href
  {\doibase 10.1103/PhysRevB.97.165129} {\bibfield  {journal} {\bibinfo
  {journal} {Phys. Rev. B}\ }\textbf {\bibinfo {volume} {97}},\ \bibinfo
  {pages} {165129} (\bibinfo {year} {2018})}\BibitemShut {NoStop}%
\bibitem [{\citenamefont {Huang}\ \emph {et~al.}(2018)\citenamefont {Huang},
  \citenamefont {Jin},\ and\ \citenamefont {Liu}}]{PhysRevLett.120.136403}%
  \BibitemOpen
  \bibfield  {author} {\bibinfo {author} {\bibfnamefont {H.}~\bibnamefont
  {Huang}}, \bibinfo {author} {\bibfnamefont {K.-H.}\ \bibnamefont {Jin}}, \
  and\ \bibinfo {author} {\bibfnamefont {F.}~\bibnamefont {Liu}},\ }\href
  {\doibase 10.1103/PhysRevLett.120.136403} {\bibfield  {journal} {\bibinfo
  {journal} {Phys. Rev. Lett.}\ }\textbf {\bibinfo {volume} {120}},\ \bibinfo
  {pages} {136403} (\bibinfo {year} {2018})}\BibitemShut {NoStop}%
\bibitem [{aps()}]{aps2018wu}%
  \BibitemOpen
  \href@noop {} {}\bibinfo {note} {Q. Wu, C. Piveteau, Z. Song, M. Troyer and
  O. V. Yazyev, ``Prediction of Dirac semimetal phase with layer-resolved
  orbital texture in ternary tantalum nitrides'', APS March meeting 2018,
  http://meetings.aps.org/Meeting/MAR18/Session/T60.22}\BibitemShut {NoStop}%
\bibitem [{pyp()}]{pyprocar}%
  \BibitemOpen
  \href@noop {} {}\bibinfo {note} {F. Munoz, and A. Romero, Pyprocar
  (2014)}\BibitemShut {NoStop}%
\bibitem [{\citenamefont {Gresch}\ \emph
  {et~al.}(2017{\natexlab{a}})\citenamefont {Gresch}, \citenamefont {Wu},
  \citenamefont {Winkler},\ and\ \citenamefont {Soluyanov}}]{Gresch2017}%
  \BibitemOpen
  \bibfield  {author} {\bibinfo {author} {\bibfnamefont {D.}~\bibnamefont
  {Gresch}}, \bibinfo {author} {\bibfnamefont {Q.}~\bibnamefont {Wu}}, \bibinfo
  {author} {\bibfnamefont {G.~W.}\ \bibnamefont {Winkler}}, \ and\ \bibinfo
  {author} {\bibfnamefont {A.~A.}\ \bibnamefont {Soluyanov}},\ }\href {\doibase
  10.1088/1367-2630/aa5de7} {\bibfield  {journal} {\bibinfo  {journal} {New
  Journal of Physics}\ }\textbf {\bibinfo {volume} {19}},\ \bibinfo {pages}
  {035001} (\bibinfo {year} {2017}{\natexlab{a}})}\BibitemShut {NoStop}%
\bibitem [{\citenamefont {Gresch}\ \emph
  {et~al.}(2017{\natexlab{b}})\citenamefont {Gresch}, \citenamefont {Wu},
  \citenamefont {Winkler},\ and\ \citenamefont
  {Soluyanov}}]{1367-2630-19-3-035001}%
  \BibitemOpen
  \bibfield  {author} {\bibinfo {author} {\bibfnamefont {D.}~\bibnamefont
  {Gresch}}, \bibinfo {author} {\bibfnamefont {Q.}~\bibnamefont {Wu}}, \bibinfo
  {author} {\bibfnamefont {G.~W.}\ \bibnamefont {Winkler}}, \ and\ \bibinfo
  {author} {\bibfnamefont {A.~A.}\ \bibnamefont {Soluyanov}},\ }\href
  {http://stacks.iop.org/1367-2630/19/i=3/a=035001} {\bibfield  {journal}
  {\bibinfo  {journal} {New Journal of Physics}\ }\textbf {\bibinfo {volume}
  {19}},\ \bibinfo {pages} {035001} (\bibinfo {year}
  {2017}{\natexlab{b}})}\BibitemShut {NoStop}%
\bibitem [{\citenamefont {Feng}\ \emph {et~al.}(2017)\citenamefont {Feng},
  \citenamefont {Yue}, \citenamefont {Song}, \citenamefont {Wu},\ and\
  \citenamefont {Wen}}]{Feng2017}%
  \BibitemOpen
  \bibfield  {author} {\bibinfo {author} {\bibfnamefont {X.}~\bibnamefont
  {Feng}}, \bibinfo {author} {\bibfnamefont {C.}~\bibnamefont {Yue}}, \bibinfo
  {author} {\bibfnamefont {Z.}~\bibnamefont {Song}}, \bibinfo {author}
  {\bibfnamefont {Q.}~\bibnamefont {Wu}}, \ and\ \bibinfo {author}
  {\bibfnamefont {B.}~\bibnamefont {Wen}},\ }\href {\doibase
  10.1103/PhysRevMaterials.2.014202} {\bibfield  {journal} {\bibinfo  {journal}
  {Physical Review Materials}\ }\textbf {\bibinfo {volume} {2}},\ \bibinfo
  {pages} {014202} (\bibinfo {year} {2017})},\ \Eprint
  {http://arxiv.org/abs/1705.00511} {arXiv:1705.00511} \BibitemShut {NoStop}%
\bibitem [{\citenamefont {Altmann}\ and\ \citenamefont
  {Herzig}(1994{\natexlab{b}})}]{altmann1994point}%
  \BibitemOpen
  \bibfield  {author} {\bibinfo {author} {\bibfnamefont {S.}~\bibnamefont
  {Altmann}}\ and\ \bibinfo {author} {\bibfnamefont {P.}~\bibnamefont
  {Herzig}},\ }\href@noop {} {\emph {\bibinfo {title} {Point-group theory
  tables}}},\ Oxford science publications\ (\bibinfo  {publisher} {Clarendon
  Press},\ \bibinfo {year} {1994})\BibitemShut {NoStop}%
\end{thebibliography}%
\bibliographystyle{apsrev4-1}

\end{document}